\documentclass[aps,11pt,prd,notitlepage,tightenlines,nofootinbib,showpacs,superscriptaddress]{revtex4-1}
\usepackage{amsmath}
\usepackage{amssymb,amsthm}
\usepackage{mathtools}
\usepackage{mathrsfs}
\usepackage{relsize}
\usepackage{bm}

\usepackage{epsfig,graphics,graphicx,color,xcolor}
\usepackage{soul} 
\usepackage{cancel} 
\usepackage{slashed}	
\usepackage{subfigure}
\usepackage{array}
\usepackage{ragged2e}
\usepackage{lineno}
\usepackage{natbib}
\usepackage{hyperref}
\hypersetup{colorlinks=true,linkcolor=red,anchorcolor=red,citecolor=orange, filecolor=brown,urlcolor=red,bookmarksnumbered=true,
pdfview=FitB
}
\graphicspath{ {image/} }
\usepackage{enumitem}

\colorlet{darkgreen}{green!50!black}
\colorlet{brightyellow}{yellow!75!red}
\colorlet{orange}{red!50!yellow}
\colorlet{darkgray}{gray!50!black}

\def\dd{{\mathrm{d}}}

\newcommand{\half}[1][1] {\mathsmaller{\frac{#1}{2}}}

\makeatletter
\newcommand*{\transpose}{%
  {\mathpalette\@transpose{}}%
}
\newcommand*{\@transpose}[2]{%
  \raisebox{\depth}{$\m@th#1\intercal$}%
}
\makeatother

\begin{document}

\title{Light cone distributions of $P$-wave quarkonia}

\author{Yang~Li}
\affiliation{Department of Modern Physics, University of Science and Technology of China, Hefei 230026, China}
\affiliation{Anhui Center for Fundamental Sciences in Theoretical Physics, University of Science and Technology of China, Hefei 230026, China}

\author{Tianyang Hu}
\affiliation{
State Key Laboratory of Heavy Ion Science and Technology, Institute of Modern Physics, Chinese Academy of Sciences, Lanzhou 730000, China}
\affiliation{School of Nuclear Science and Technology, University of Chinese Academy of Sciences, Beijing 100049, China}

\author{Xianghui Cao}
\affiliation{Department of Modern Physics, University of Science and Technology of China, Hefei 230026, China}

\author{Meijian Li}
\thanks{Corresponding author: meijianli@hust.edu.cn}
\affiliation{School of Physics, Huazhong University of Science and Technology, Wuhan 430074, China}

\date{\today}

\begin{abstract}
Motivated by renewed interest in the light-cone distributions of $P$-wave quarkonia, we investigate the leading-twist distribution amplitudes of these states within the basis light-front quantization (BLFQ) formalism. While our extracted distributions exhibit macroscopic shapes consistent with expectations from the non-relativistic limit, we find that relativistic effects introduce critical structural features. Most notably, we observe novel ``W"-shaped structures in the distribution amplitudes of the axial vector mesons, which arise from relativistically induced $S/D$ partial waves and cannot be explained by non-relativistic dynamics. These findings provide essential non-perturbative inputs for analyzing hard exclusive processes and highlight the importance of relativistic frameworks for understanding heavy quarkonium production and structure at modern high-energy colliders.
\end{abstract}
\maketitle 

\section{Introduction}

Light cone distributions encode the non-perturbative structural information of hadrons, and are indispensable tools for describing modern high-energy scattering experiments.  In particular, parton distribution functions (PDFs) and light-cone distribution amplitudes (LCDAs) control the inclusive and exclusive processes at large momentum transfer, respectively \cite{Lepage:1980fj}. 
Recently, there has been renewed interest in the light-cone distributions of $P$-wave quarkonia \cite{Yang:2005gk, Yang:2007zt, Hwang:2009cu, Verma:2011yw, Hwang:2012nw, Wang:2013ywc, Olpak:2016wkf, Li:2016mah, Rui:2017pre, Rui:2018kqr, Sungu:2018eej,  Aliev:2017apq, Hoferichter:2020lap, Chen:2021vmb, Akan:2022vtf, Zhang:2023ypl, Akan:2025mfr, Lu:2025bvi, Liu:2025ipe, Zeng:2025gft, Zhang:2026zmu, Zeng:2026peb, Xu:2026zli}. Experimentally, the abundant production of $\chi_{cJ}$ states at $e^+e^-$ and hadron-hadron colliders, such as BESIII \cite{BESIII:2012uyb, BESIII:2012mpj, BESIII:2015som, Belle:2015hcs, BESIII:2016hfo, BESIII:2017ung, BESIII:2017tsq, BESIII:2021yal, BESIII:2022mtl, BESIII:2024jmr, BESIII:2025umc}, Belle \cite{Belle:2002tfa, Belle:2002yet, Belle:2015opn, Belle:2022exn}, and LHC \cite{LHCb:2013ofo, ATLAS:2014ala, LHCb:2014ngh, LHCb:2014nug, LHCb:2017hzb, CMS:2018vgf}, provides excellent opportunities to investigate their properties and structures. Theoretically, $P$-wave mesons represent angular excited states. Because spin ($S$) and orbital ($L$) angular momenta are not strictly conserved quantum numbers in a relativistic framework, it is highly instructive to examine how angular excitation interplays with relativistic dynamics to dictate the shape of $P$-wave mesons. Charmonia and bottomonia serve as ideal testbeds for these studies, as they straddle the boundary between relativistic and non-relativistic regimes, facilitating clear comparisons across different theoretical methods \cite{Gross:2022hyw, Brambilla:2010cs}.

Theoretical access to these quantities has proven challenging, especially for theories formulated in Euclidean spacetime, e.g., Lattice QCD. This difficulty arises because light-cone distributions are intrinsically \emph{Minkowskian}, despite major advances in recent years bridging Euclidean and Minkowski spacetimes \cite{Karmanov:2005nv, Ji:2013dva, Chang:2013pq, Shi:2018zqd, Ji:2020ect}.
Alternatively, light-front wave functions (LFWFs) provide more direct access to these distributions \cite{Brodsky:1997de}. LFWFs are the eigenfunctions of the light-cone Hamiltonian $H_\textsc{lc} = P^+P^--\vec P_\perp^2$, where $P^\pm = P^0\pm P^3$ and $\vec P_\perp = (P^1, P^2)$. As frame-independent quantities, they offer a unified framework for computing hadronic observables, including form factors, decay constants, and light-cone distributions \cite{Vary:2025yqo, Shuryak:2026pqt}.

In this work, we present the leading-twist LCDAs of $P$-wave quarkonia using an effective light-cone Hamiltonian developed within the basis light-front quantization (BLFQ) framework \cite{Vary:2009gt}. The underlying interaction combines light-front holographic QCD confinement with a short-distance one-gluon exchange, which is essential for generating proper hadronic spin structures \cite{Li:2015zda, Li:2017mlw}. This model relies on two phenomenological parameters, i.e. the quark mass and confining strength, determined by fitting to experimental spectra. When applied to heavy quarkonia, the resulting mass spectra, decay constants, and form factors agree well with both experimental measurements and alternative theoretical approaches \cite{Li:2017mlw}. This framework has also been extended to other heavy mesons ($B_c$, $B$, $D$, $B_s$, and $D_s$) and, with modifications, to light mesons \cite{Tang:2018myz, Tang:2019gvn, Qian:2020utg}. Building on our previous computation of $S$-wave meson light-cone distributions \cite{Li:2017mlw}, where the extracted moments aligned with QCD sum rules, Dyson-Schwinger equations, NRQCD, and Covariant Spectator Theory, we now focus exclusively on $P$-wave states.

This paper is organized as follows. In Section \ref{sec:P-wave}, we investigate the covariant structures of the LFWFs and obtain the corresponding LCDAs from them. Section \ref{sec:numerical_results} provides a brief overview of the theoretical formalism before detailing the numerical results. Finally, we conclude in Section \ref{sec:summary}.

\section{$P$-wave quarkonia}\label{sec:P-wave}

The primary $P$-wave quarkonia of interest are the scalar $S$ ($0^{++}$), axial vectors $A$ ($1^{++}$) and $h$ ($1^{+-}$), and the tensor $T$ ($2^{++}$).
In the \textit{non-relativistic} quark model, these states are mapped to the ${}^3P_0$, ${}^3P_1$, ${}^1P_1$, and ${}^3P_2$ configurations, respectively, each characterized by an orbital angular momentum $L=1$. However, relativistic spin-orbit coupling mixes states with different orbital angular momenta $L$ and total spins $S$, meaning these are no longer strictly conserved quantum numbers. This mixing is constrained by angular momentum addition, $\vert{}L-S\vert{} \le J \le L+S$, alongside parity and charge conjugation symmetries: $\textsf{P} = (-1)^{L+1}$ and $\textsf{C} = (-1)^{L+S}$. Within the valence sector, these constraints dictate that partial-wave mixing occurs only for the tensor meson $T$ ($2^{++}$), which blends the ${}^3P_2$ and ${}^3F_2$ configurations. For the remaining $P$-wave states, any partial-wave mixing requires higher Fock components beyond the valence $q\bar{q}$ sector.

Relativistic dynamics fundamentally alter the partial-wave composition of hadrons. While mesons are traditionally classified by discrete quantum numbers like parity $\textsf{P}$ and charge conjugation $\textsf{C}$, imposing these symmetries in light-front dynamics requires careful treatment. Because the standard parity operator exchanges the light-front time ($x^+$) with the longitudinal spatial coordinate ($x^-$), it acts as a dynamical operator rather than a manifest kinematical symmetry. A convenient, kinematic alternative is mirror parity, $m_\textsf{P}: (x, y, z) \to (-x, y, z)$, which reflects only a single spatial transverse coordinate. This discrete symmetry remains strictly conserved on the light front and relates to standard parity via $m_\textsf{P}=(-i)^{2J}\textsf{P}$, where $J$ is the total angular momentum \cite{Brodsky:2006ez}. These relativistic symmetry constraints explicitly govern the generation of new partial waves. 

For instance, the pseudoscalar meson $P$ ($0^{-+}$) is a purely ${}^1S_0$ state in the non-relativistic quark model. 
In light-front dynamics, the exact mirror parity must be $m_\textsf{P} = -1$. Under the $m_\textsf{P}$ operation, the transverse momentum reflects $k_x \to -k_x$ and the constituent helicities flip $\uparrow\uparrow \; \leftrightarrow \;\downarrow\downarrow$. To satisfy $m_\textsf{P} = -1$ while maintaining a total longitudinal angular momentum projection of $J_z = 0$, any spin-aligned configuration must be intrinsically coupled to a compensating unit of orbital angular momentum $| L_z | = 1$. This geometric symmetry requirement mandates a spatial dependence proportional to the transverse momentum phase $(k_x \pm ik_y)$. Consequently, the pseudoscalar meson acquires an explicit spin-flip $P$-wave component, $\psi_{\uparrow\uparrow/P}(x, \vec k_\perp) \propto (k_x+ik_y)\phi(x, k_\perp)$, not as a violation of parity, but exactly as a necessary consequence of enforcing mirror parity on the light front.

A $P$-wave component also emerges in the Bethe-Salpeter amplitude (BSA) of pseudoscalar mesons evaluated in the rest frame \cite{Maris:2003vk}. However, it arises from entirely different Lorentz structures, namely $\gamma_5\slashed{k}$ and $\gamma_5\sigma_{\mu\nu} k^\mu P^\nu$, where $k=p-\bar p$ and $P=p+\bar p$ are the relative and total meson momenta, respectively, and $p$ and $\bar p$ are the 4-momenta of the quark and antiquark, respectively.  Crucially, the $\gamma_5\slashed{k}$ term vanishes in light-front dynamics because the constituent quark's momenta $p$ and $\bar p$ are placed on their mass shells; applying the Dirac equation for equal-mass quarks ($m_q = m_{\bar{q}}$) yields exactly zero for this matrix element.
Physically, a 3D reduction of the BSA explicitly depends on the chosen reference frame (such as the meson rest frame), whereas LFWFs are explicitly invariant under kinematic light-front boosts. Because there is no general one-to-one mapping between the invariant Lorentz structures of DSEs/BSEs and those of light-front dynamics, structural comparisons between these approaches must be performed in a common kinematic limit. The most rigorous method is to match them in the infinite momentum frame by comparing their respective LCDAs \cite{Cao:2025bit}.

A systematic way to implement these symmetries on the light front is the covariant light-front dynamics (CLFD) \cite{Carbonell:1998rj}. In this framework, the pseudoscalar LFWF decomposes as:
\begin{equation}
\psi_{s\bar s/P}(x, \vec k_\perp) = \bar u_s(p)\left[
\gamma_5 \phi_1(x, k_\perp) + \frac{M_P \gamma^+\gamma_5}{P^+} \phi_2(x, k_\perp) \right] v_{\bar s}(\bar p),
\end{equation}
where $p$, $\bar p$, and $P$ are the 4-momenta of the quark, antiquark, and meson, respectively. In standard light-front coordinates $(p^+, \vec{p}_\perp, p^-)$, the on-shell constituents carry momenta $p = \big(xP^+, \vec k_\perp+x\vec P_\perp, \frac{(\vec k_\perp+x\vec P_\perp)^2+m^2_q}{xP^+}\big)$ and $\bar p = \big((1-x)P^+, -\vec k_\perp+(1-x)\vec P_\perp, \frac{(-\vec k_\perp+(1-x)\vec P_\perp)^2+m^2_{\bar q}}{(1-x)P^+}\big)$. And $k_\perp = |\vec k_\perp|$ is the magnitude. Evaluating the spinor matrix elements explicitly yields the spin-flip $P$-wave term: $\psi_{\downarrow\downarrow/P}(x, \vec k_\perp) = \psi_{\uparrow\uparrow/P}^*(x, \vec k_\perp) = - (k_x+ik_y)\phi_1(x, k_\perp)/\sqrt{x(1-x)}$.  
Crucially, this CLFD decomposition represents the most general structural form permitted by exact symmetries, extending well beyond traditional ansätze based solely on the Melosh rotation \cite{Choi:1996mq, Jaus:1999zv}. While the Melosh (or light-front Wigner) rotation correctly maps canonical rest-frame spins to light-front helicities, it only generates standard manifestly covariant structures, such as the pure $\gamma_5$ term \cite{Melosh:1974cu}. It fundamentally misses dynamics tied to the choice of the quantization axis. By systematically incorporating these explicit light-front orientation-dependent terms, such as $(\gamma^+\gamma_5)/P^+$, CLFD ensures the strict restoration of full Lorentz covariance in physical observables. 

Below, we derive the general covariant light-front structures of $P$-wave quarkonia and formulate their leading-twist LCDAs in terms of the LFWFs. The light-front null vector that dictates the orientation of light-front time is denoted as $\omega^\mu$ ($\omega_\mu \omega^\mu = 0$, $\omega _\mu a^\mu = a^+$). In standard light-front coordinates $x^\pm = x^0 \pm x^3$, it takes the form $\omega^\mu = (\omega^0, \vec \omega) = (1, 0, 0, -1)$.  It is convenient to construct a scaled null vector: $\omega^\mu/(\omega \cdot P)$ to incorporate the manifest invariance under longitudinal scaling $\omega^\mu \to \lambda \omega^\mu$. 

\subsection{Scalar $\chi_{0}$ ($0^{++}$) }

\begin{figure}
    \centering
    \subfigure[\ $\psi_{\uparrow\downarrow+\downarrow\uparrow}$]{\includegraphics[width=0.3\textwidth]{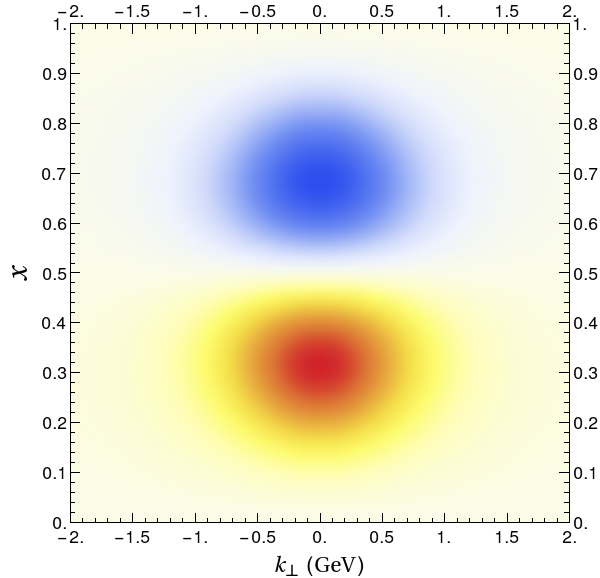}} \qquad
    \subfigure[\ $\psi_{\downarrow\downarrow}$]{\includegraphics[width=0.3\textwidth]{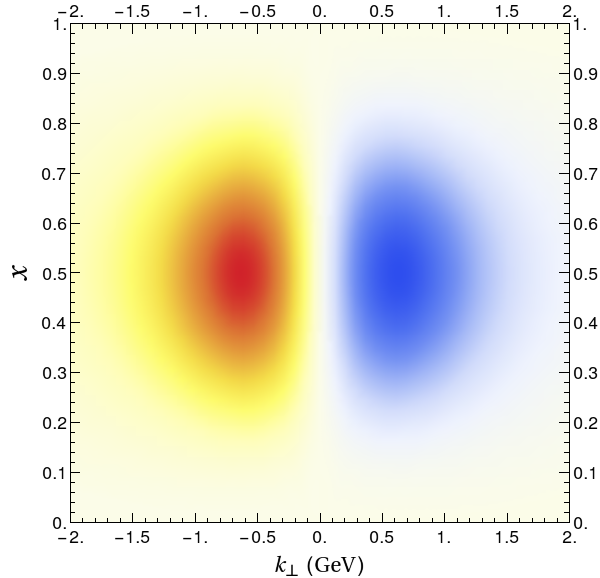}}
    \caption{LFWFs of scalar quarkonium $\chi_{c0}(1P)$ obtained from BLFQ. }
    \label{fig:Sc}
\end{figure}

The general covariant light-front structure of a scalar quarkonium ($S$) is parameterized by two invariant functions, $\phi_1$ and $\phi_2$:
\begin{equation}
\psi_{s\bar s/S}(x, \vec k_\perp) = \bar u_s(p)\left[
\phi_1(x, k_\perp) + \frac{M_S \gamma^+}{P^+} \phi_2(x, k_\perp) \right] v_{\bar s}(\bar p).
\end{equation}
Evaluating the spinor matrix elements yields two independent spin configurations:
\begin{equation}
\begin{split}
    & \psi_{\uparrow\uparrow/S}(x, \vec k_\perp) = - \psi^*_{\downarrow\downarrow/S}(x, \vec k_\perp) = \frac{k^1_\perp- ik^2_\perp} {\sqrt{x(1-x)}} \phi_1(x, k_\perp), \\
    & \psi_{\uparrow\downarrow+\downarrow\uparrow/S}(x, \vec k_\perp) = \sqrt{2} m_q \frac{1-2x}{\sqrt{x(1-x)}} \phi_1(x, k_\perp) + 2\sqrt{2} M_S\sqrt{x(1-x)}\phi_2(x, k_\perp)\;,
\end{split}
\end{equation} 
where we adopt the normalized shorthand notation $\psi_{\uparrow\downarrow\pm\downarrow\uparrow}  = \big[\psi_{\uparrow\downarrow}  \pm \psi_{\downarrow\uparrow}  \big]/\sqrt{2}$. Because the scalar meson is strictly $\textsf{C}$-even ($+1$), charge conjugation symmetry constrains the invariant functions under longitudinal momentum exchange: $\phi_1(x, k_\perp)$ must be symmetric $\phi_1(1-x, k_\perp) = \phi_1(x, k_\perp)$, while $\phi_2(x, k_\perp)$ must be antisymmetric $\phi_2(1-x, k_\perp) = -\phi_2(x, k_\perp)$.

Figure~\ref{fig:Sc} displays the LFWFs of the scalar charmonium $\chi_{c0}(1P)$ computed using the BLFQ approach. The distributions exhibit distinct geometric nodes characteristic of $P$-wave states. The $\psi_{\uparrow\uparrow/S}$ and $\psi_{\downarrow\downarrow/S}$ components vanish at the transverse origin ($k_\perp = 0$), while $\psi_{\uparrow\downarrow+\downarrow\uparrow/S}$ features a longitudinal node at $x=1/2$ induced by the $1-2x$ factor. This explicitly mirrors the spin-momentum correlations found in the non-relativistic ${}^3P_0$ wave function:
\begin{equation}
\Psi_{s\bar s/S}(\vec k) = ({{1}/{\sqrt{8\pi}}}) \big[\hat k\cdot\vec\sigma\sigma_2\big]_{s\bar s} \phi_S(|\vec k|),
\end{equation}
where $\sigma_i$ are the Pauli matrices.

The leading-twist (twist-2) LCDA is extracted from the light-like separated vector correlator:
\begin{equation}\label{eqn:S}
     \langle0|\overline\psi(-z)\gamma^+\psi(+z)|S(p)\rangle\big|_{z^+=z_\perp=0}
= p^+f_S \int_0^1 \dd x \, \exp\big[ i (x-\half) p^+z^- \big] \phi_S(x)\;.
\end{equation}
Because the vector current $\overline\psi\gamma^+\psi$ is $\textsf C$-odd and the scalar meson is $\textsf C$-even, the local matrix element strictly vanishes $\langle0\vert{}\overline\psi(0)\gamma^+\psi(0)\vert{}S(p)\rangle = 0$. Consequently, the normalization constant $f_S$ is not a conventional decay constant, and the zeroth moment of the LCDA is zero. However, in the non-relativistic limit, $f_S$ correlates directly with the decay constant of axial-vector states of the same radial excitation \cite{Braguta:2008qe}. To account for this, the distribution amplitude $\phi_S$ is normalized to its first moment rather than its zeroth:
\begin{equation}
\int_0^1 \dd x (2x-1) \phi_S(x) = 1.
\end{equation}
Projecting this definition onto the LFWFs provides the explicit LCDA representation \cite{Lepage:1980fj}:
\begin{equation}\label{eqn:LCDA_S}
\frac{f_S}{2\sqrt{2N_c}} \phi_S(x; \mu) =  \frac{1}{\sqrt{x(1-x)}} \int^{\mu^2} \frac{\dd^2 k_\perp}{2(2\pi)^3}
\psi_{\uparrow\downarrow+\downarrow\uparrow/S}(x, \vec k_\perp).
\end{equation}
In the BLFQ framework, the finite basis size inherently imposes an ultraviolet cutoff, tying the factorization scale directly to the basis truncation: $\mu = \Lambda_{\text{UV}} = \kappa \sqrt{N_{\text{max}}}$. Explicit dependence on the scale $\mu$ is implicitly understood in subsequent expressions.

\subsection{Axial vector $\chi_{1}$ ($1^{++}$) }

\begin{figure}
    \centering
    \subfigure[\ $\psi_{\uparrow\downarrow-\downarrow\uparrow}^{(m_j=0)}$]{\includegraphics[width=0.3\textwidth]{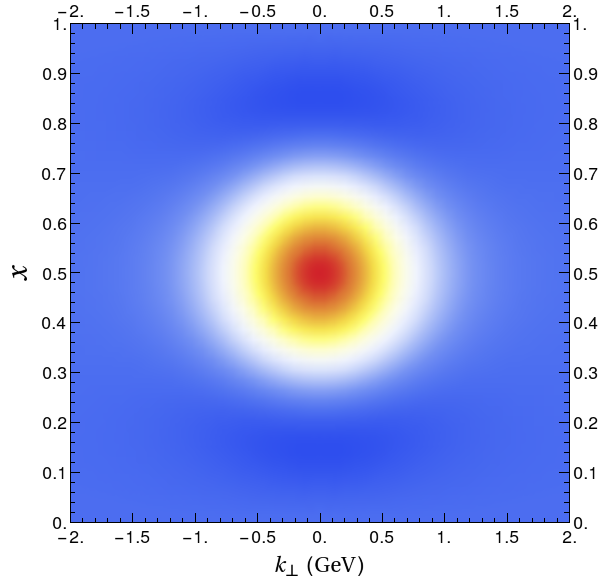}} \;
    \subfigure[\ $\psi_{\downarrow\downarrow}^{(m_j=0)}$]{\includegraphics[width=0.3\textwidth]{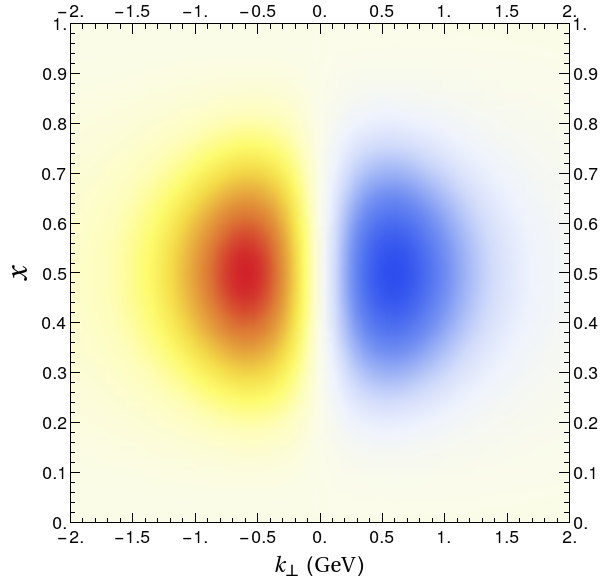}} \;
    \subfigure[\ $\psi_{\uparrow\downarrow-\downarrow\uparrow}^{(m_j=1)}$]{\includegraphics[width=0.3\textwidth]{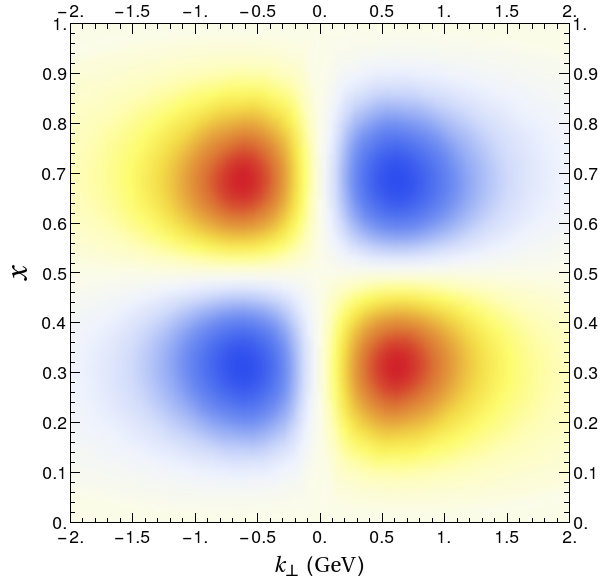}} \\
    \subfigure[\ $\psi_{\uparrow\downarrow+\downarrow\uparrow}^{(m_j=1)}$]{\includegraphics[width=0.3\textwidth]{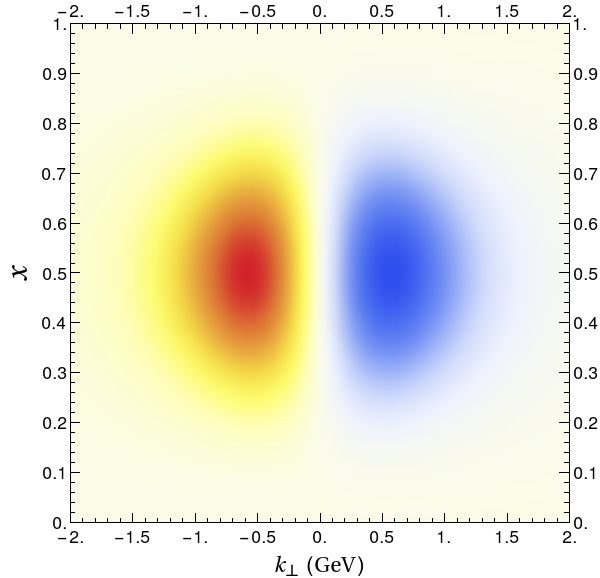}} \;
    \subfigure[\ $\psi_{\uparrow\uparrow}^{(m_j=1)}$]{\includegraphics[width=0.3\textwidth]{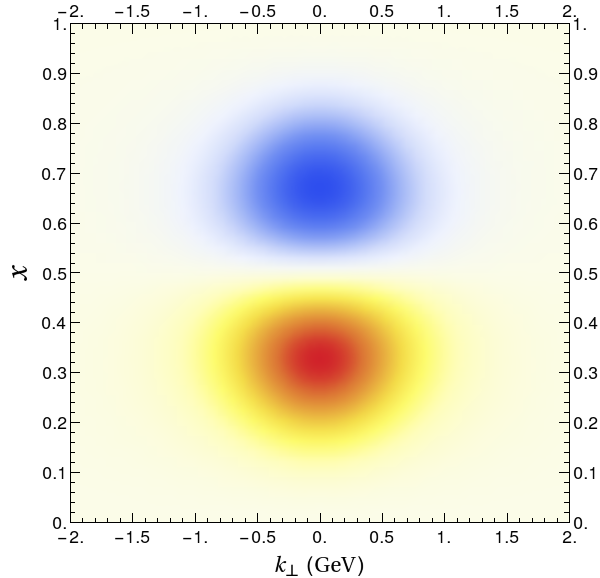}} \;
    \subfigure[\ $\psi_{\downarrow\downarrow}^{(m_j=1)}$]{\includegraphics[width=0.3\textwidth]{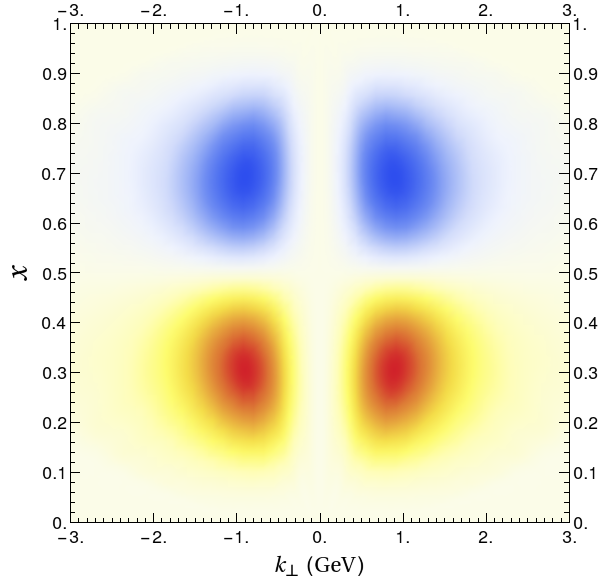}} 
    \caption{Light-front wave functions of axial vector quarkonium $\chi_{c1}(1P)$ obtained from BLFQ. }
    \label{fig:Ac}
\end{figure}

The general covariant light-front structure of an axial vector meson ($J^P = 1^+$) is parameterized by six invariant functions, $\phi_1$ through $\phi_6$:
\begin{multline}\label{eqn:CLFQ_axial_vector}
\psi_{s\overline{s}/A}^{\lambda}(x,\vec{k}_{\perp}) = e_{\mu}^{\lambda}(P)\overline{u}_{s}(p)\bigg[ (p-\overline{p})^{\mu}\gamma_5\phi_{1}(x,k_{\perp}) + \gamma^{\mu}\gamma_5\phi_{2}(x,k_{\perp}) + \frac{\omega^{\mu}}{P^{+}}\gamma_5\phi_{3}(x,k_{\perp})  \\
+ \frac{\gamma^{+}}{P^{+}}(p-\overline{p})^{\mu}\gamma_5\phi_{4}(x,k_{\perp}) + i\epsilon^{\mu\nu\rho\sigma} P_{\nu}(p-\overline{p})_{\rho}\frac{\omega_{\sigma}}{P^{+}}\phi_{5}(x,k_{\perp})  
+ \frac{\omega^{\mu}\gamma^{+}\gamma_5}{(P^{+})^{2}}\phi_{6}(x,k_{\perp}) \bigg] v_{\overline{s}}(\overline{p})\;,
\end{multline}
where the dependence of $\phi_i(x, k_\perp)$ on the kinematic variables is implicitly understood. This Lorentz structure universally accommodates both the $1^{++}$ and $1^{+-}$ states, with the distinction enforced by charge conjugation symmetry $\textsf{C}$. Under longitudinal momentum fraction exchange ($x \leftrightarrow 1-x$), the invariant functions for the $1^{++}$ ($\textsf{C}=+1$) state obey distinct parity rules: $\phi_{1,4,5}$ are antisymmetric (odd), while $\phi_{2,3,6}$ are symmetric (even). Conversely, for the pseudovector $1^{+-}$ ($\textsf{C}=-1$) state, $\phi_{1,4,5}$ are symmetric and $\phi_{2,3,6}$ are antisymmetric.

Projecting out the spinor matrix elements yields the explicit partial-wave spin configurations. For the longitudinal polarization ($\lambda = 0$):
\begin{multline}
\psi_{\uparrow\uparrow/A}^{\lambda=0}(x,\vec{k}_{\perp})=\psi_{\downarrow\downarrow/A}^{\lambda=0{*}}(x,\vec{k}_{\perp})=\frac{k_{\perp}^{1}-ik_{\perp}^{2}}{2M_{A}}\frac{2x-1}{\sqrt{x(1-x)}}\left[\frac{k_{\perp}^{2}+m_{q}^{2}}{x(1-x)}+M_{A}^{2}\right]\phi_{1}(x,k_{\perp}) \\
- \frac{k_{\perp}^{1}-ik_{\perp}^{2}}{M_{A}}\frac{2m_q}{\sqrt{x(1-x)}}\phi_{2}(x,k_{\perp})-\frac{k_{\perp}^{1}-ik_{\perp}^{2}}{M_{A}}\frac{1}{\sqrt{x(1-x)}}\phi_{3}(x,k_{\perp})\;,
\end{multline}
\begin{multline}
\psi_{\uparrow\downarrow-\downarrow\uparrow/A}^{\lambda=0}(x,\vec{k}_{\perp})=-\frac{m_q}{\sqrt{2}M_{A}}\frac{2x-1}{\sqrt{x(1-x)}}\left[\frac{k_{\perp}^{2}+m^{2}_q}{x(1-x)}+M_{A}^{2}\right]\phi_{1}(x,k_{\perp}) \\
- \frac{\sqrt{2x(1-x)}}{M_{A}}\left[\frac{{k_{\perp}^{2} - m^{2}_q}}{x(1-x)}+M_{A}^{2}\right]\phi_{2}(x,k_{\perp})+\frac{\sqrt{2}m_q}{M_{A}}\frac{1}{\sqrt{x(1-x)}}\phi_{3}(x,k_{\perp}) \\
- \frac{\sqrt{2}x(1-x)}{M_{A}}\frac{2x-1}{\sqrt{x(1-x)}}\left[\frac{k_{\perp}^{2}+m^{2}_q}{x(1-x)}+M_{A}^{2}\right]\phi_{4}(x,k_{\perp})+\frac{2\sqrt{2x(1-x)}}{M_{A}}\phi_{6}(x,k_{\perp})\;.
\end{multline}

For the transverse polarizations ($\lambda = +1$):
\begin{equation}
\psi_{\uparrow\uparrow/A}^{\lambda=+1}(x,\vec{k}_{\perp})=\frac{\sqrt{2}\vec k_{\perp}^{2}}{\sqrt{x(1-x)}}\phi_{1}(x,k_{\perp})+\frac{\sqrt{2}m_q(2x-1)}{\sqrt{x(1-x)}}\phi_{2}(x,k_{\perp})-\frac{\sqrt{2}\vec k_{\perp}^{2}}{\sqrt{x(1-x)}}\phi_{5}(x,k_{\perp})\;,
\end{equation}
\begin{equation}
\psi_{\uparrow\downarrow+\downarrow\uparrow/A}^{\lambda=+1}(x,\vec{k}_{\perp})= \frac{k_{\perp}^{1}+ik_{\perp}^{2}}{\sqrt{x(1-x)}}\phi_{2}(x, \vec k_{\perp}) +2m_q(k_{\perp}^{1}+ik_{\perp}^{2})\frac{2x-1}{\sqrt{x(1-x)}}\phi_{5}(x,  k_{\perp})\;,
\end{equation}
\begin{multline}
\psi_{\uparrow\downarrow-\downarrow\uparrow/A}^{\lambda=+1}(x,\vec{k}_{\perp}) = 
-\frac{2m_q(k^1_\perp+i k^2_\perp)}{\sqrt{x(1-x)}}\phi_1(x, k_\perp) + \frac{(2x-1)(k^1_\perp+i k^2_\perp)}{\sqrt{x(1-x)}} \phi_2(x, k_\perp) \\
- 4 \sqrt{x(1-x)}(k^1_\perp+ik^2_\perp) \phi_4(x, k_\perp)\;,
\end{multline}
\begin{equation}
    \psi_{\downarrow\downarrow / A}^{\lambda = +1}(x, \vec{k}_\perp) = \sqrt{\frac{2}{x(1-x)}} (k_\perp^1 + i k_\perp^2)^2 \Big[ \phi_1(x, k_\perp) + \phi_5(x, k_\perp) \Big]\;.
\end{equation}

Figure~\ref{fig:Ac} displays the LFWFs of the axial vector meson $\chi_{c1}(1P)$ ($1^{++}$). Crucially, relativistic spin-orbit coupling on the light front generates partial waves beyond the standard $P$-wave configurations. We observe distinct $S/D$-wave components $\psi_{\uparrow\downarrow-\downarrow\uparrow}^{(m_j=0)}$ and $\psi_{\uparrow\downarrow-\downarrow\uparrow}^{(m_j=+1)}$ and an $F$-wave component $\psi_{\downarrow\downarrow}^{(m_j=+1)}$. Among these dynamically generated configurations, the $S/D$-wave state $\psi_{\uparrow\downarrow-\downarrow\uparrow}^{(m_j=0)}$ is particularly robust and directly drives the leading-twist distribution amplitude. Similar mixed partial waves are present in the LFWFs of the bottomonium equivalent, $\chi_{b1}(1P)$. These relativistic findings contrast sharply with the standard non-relativistic ${}^3P_1$ wave function:
\begin{equation}
\Psi_{s\bar s/A}^{(m_j)}(\vec k) = \frac{3}{4\sqrt{\pi}} \big[\vec \sigma \cdot (\vec \xi^{m_j} \times \hat k) \sigma_2\big]_{s\bar s} \phi_A(|\vec k|),
\end{equation}
where $\vec \xi^\pm=(1,\pm i,0)/\sqrt{2}$ and $\vec \xi^0=(0,0,1)$ represent the polarization vectors. While the $S/D$- and $F$-wave spin components are absent in the non-relativistic limit, they are strictly allowed by discrete symmetries, specifically mirror parity $m_\textsf{P}$ and charge conjugation $\textsf{C}$.

The longitudinal leading-twist LCDA of the axial vector quarkonia is defined via the light-like correlator\footnote{In the literature, $\phi_A$ is frequently denoted as $\phi_A^\Vert{}$, to explicitly contrast with the transverse $\phi_A^\perp$.}:
\begin{equation}\label{eqn:A}
    \langle0|\overline\psi(-z)\gamma^+\gamma_5\psi(+z)|A(p, \lambda=0)\rangle\big|_{z^+=z_\perp=0}
=\, p^+f_A \int_0^1 \dd x \, \exp\big[ i (x-\half) p^+z^- \big] \phi_A(x),  
\end{equation}
where $f_A$ is the decay constant, determined by the local matrix element $\langle0|\overline\psi(0)\gamma^+\gamma_5\psi(0)|A(p, \lambda=0)\rangle
= p^+f_A$. This establishes the unit normalization condition:
\begin{equation}
    \int_0^1 \dd x \, \phi_A(x) = 1.
\end{equation}
Projecting this onto the LFWF basis yields:
\begin{equation}\label{eqn:LCDA_A}
    \frac{f_A}{2\sqrt{2N_c}} \phi_A(x) =  \frac{1}{\sqrt{x(1-x)}} \int \frac{\dd^2 k_\perp}{2(2\pi)^3}
    \psi^{(m_j=0)}_{\uparrow\downarrow-\downarrow\uparrow/A}(x, \vec k_\perp).
\end{equation}
Note that the longitudinal DA $\phi_A(x)$ is entirely determined by the $S/D$-wave LFWF $\psi^{(m_j=0)}_{\uparrow\downarrow-\downarrow\uparrow/A}$, which underscores the necessity of the relativistic framework, as this component strictly vanishes in the non-relativistic limit.

A second leading-twist LCDA governs the transverse dynamics and isolates the $L_z=0$ LFWF, $\psi_{\uparrow\uparrow}^{(m_j=1)}$. It is accessed via the tensor current correlator:
\begin{equation}
    \langle 0|\overline\psi(-z)\sigma^{+i} \psi(+z)|A(p, \lambda=\pm1)\rangle\big|_{z^+=z_\perp=0} = p^+\epsilon^{ij}e^j_{\pm 1}(p) f_A^{\perp} \int_0^1 \dd x\,\exp\big[i(x-\half)p^+z^-\big]\phi_{A}^\perp(x),
\end{equation}
where $\sigma^{\mu\nu}=\frac{i}{2}\big[\gamma^\mu,\gamma^\nu\big]$ and $\epsilon^{ij}$ ($i,j \in \{1,2\}$) is the fully antisymmetric transverse tensor. The constant $f_A^\perp$ normalizes the transverse LCDA to its first moment:\begin{equation}
    \int_0^1 \dd x\, (2x-1)\phi_A^\perp(x) = 1. 
\end{equation}
The LFWF representation of the transverse distribution amplitude $\phi_A^\perp(x)$ is therefore given by:
\begin{equation}\label{eqn:LCDA_Aperp}
\frac{f_A^\perp}{2\sqrt{2N_c}}\phi_A^\perp(x) = \frac{1}{\sqrt{x(1-x)}}\int \frac{\dd^2k_\perp}{2(2\pi)^3} \psi_{\uparrow\uparrow/A}^{(m_j=+1)}(x, \vec k_\perp).
\end{equation}

\subsection{Axial vector $h$ ($1^{+-}$)}

\begin{figure}
    \centering
    \subfigure[\ $\psi_{\uparrow\downarrow-\downarrow\uparrow}^{(m_j=0)}$]{\includegraphics[width=0.3\textwidth]{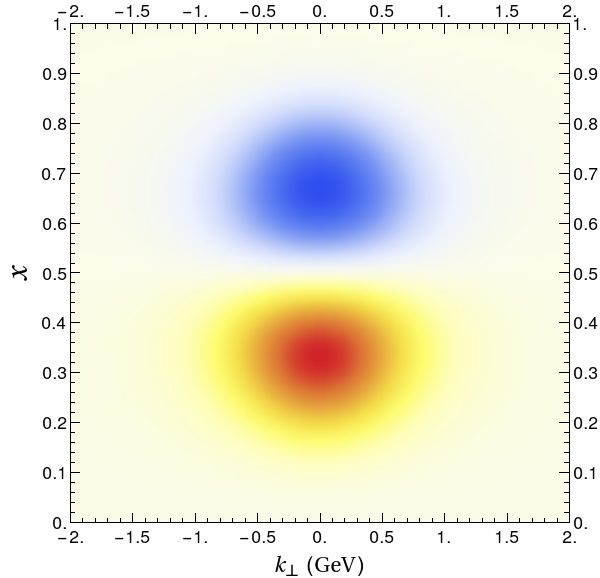}} \;
    \subfigure[\ $\psi_{\downarrow\downarrow}^{(m_j=0)}$]{\includegraphics[width=0.3\textwidth]{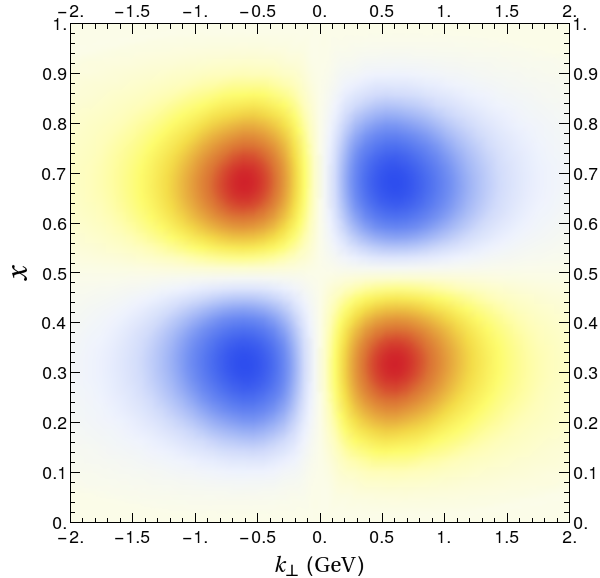}} \;
    \subfigure[\ $\psi_{\uparrow\downarrow-\downarrow\uparrow}^{(m_j=1)}$]{\includegraphics[width=0.3\textwidth]{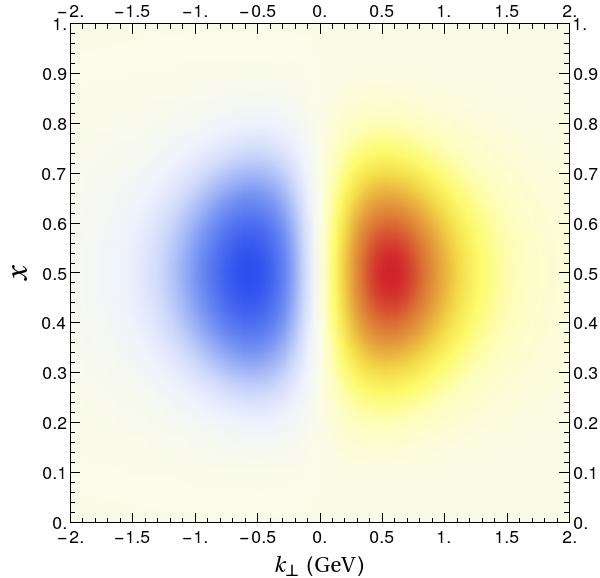}} \\
    \subfigure[\ $\psi_{\uparrow\downarrow+\downarrow\uparrow}^{(m_j=1)}$]{\includegraphics[width=0.3\textwidth]{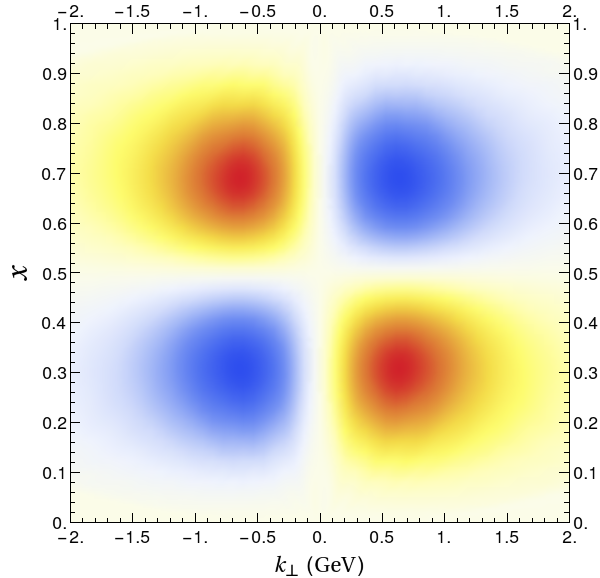}} \;
    \subfigure[\ $\psi_{\uparrow\uparrow}^{(m_j=1)}$]{\includegraphics[width=0.3\textwidth]{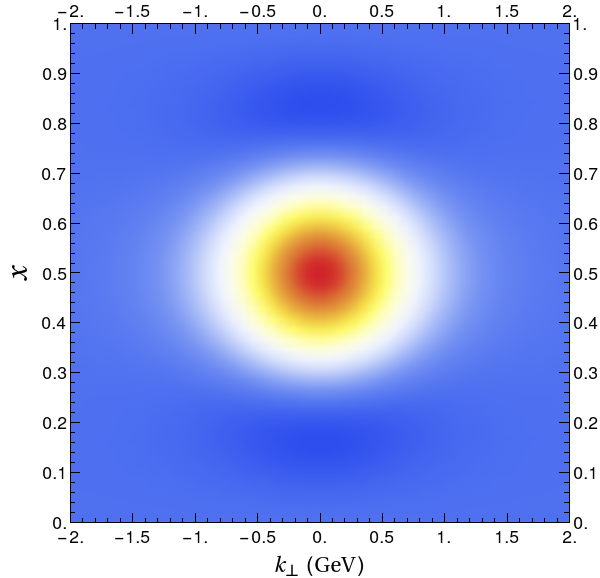}} \;
    \subfigure[\ $\psi_{\downarrow\downarrow}^{(m_j=1)}$]{\includegraphics[width=0.3\textwidth]{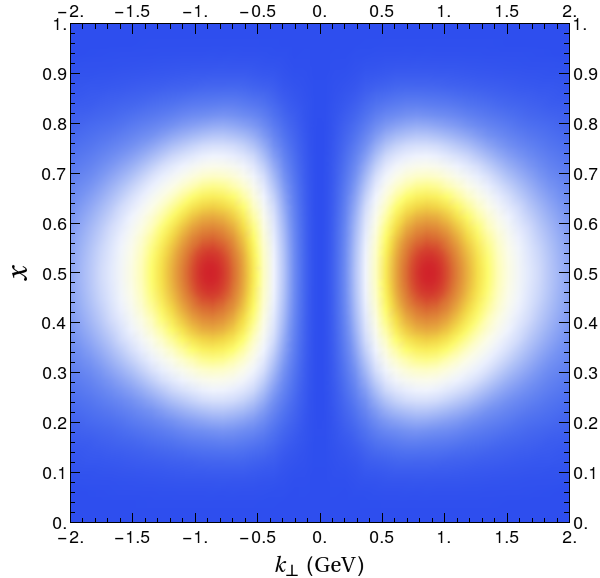}} 
    \caption{Light-front wave functions of axial vector quarkonium $h_{c}(1P)$ obtained from BLFQ. }
    \label{fig:hc}
\end{figure}

The $h_c$ and $h_b$ mesons represent the second class of axial vector quarkonia ($J^{PC} = 1^{+-}$), distinguished from the $\chi$ states by a negative charge conjugation symmetry $\textsf{C} = -1$. Their general covariant light-front structure is parameterized by the identical Lorentz decomposition given in Eq.~(\ref{eqn:CLFQ_axial_vector}). However, the $\textsf{C} = -1$ constraint reverses the symmetry requirements of the invariant functions under longitudinal momentum exchange ($x \leftrightarrow 1-x$): the components $\phi_{1,4,5}$ must be symmetric (even), while $\phi_{2,3,6}$ must be antisymmetric (odd). Figure~\ref{fig:hc} illustrates the LFWFs of the $h_c(1P)$ state. In the non-relativistic limit, these mesons are pure ${}^1P_1$ states characterized by an orbital angular momentum $L=1$ and total spin $S=0$. The non-relativistic wave function naturally takes a purely spin-singlet form:
\begin{equation}
\Psi_{s\bar s/h}^{(m_j)}(\vec k) = \frac{3}{\sqrt{8\pi}}\big[\vec\xi^{m_j}\cdot\hat k\sigma_2\big]_{s\bar s}\phi_h(|\vec k|).
\end{equation}
However, on the light front, relativistic spin-orbit interactions dynamically break heavy quark spin symmetry. Consequently, the LFWFs develop spin-triplet ($S=1$) configurations alongside the expected spin-singlet $P$-wave states. Specifically, we observe the emergence of structurally distinct $S/D$-wave components, including $\psi_{\uparrow\uparrow}^{(m_j=1)}$, $\psi_{\downarrow\downarrow}^{(m_j=0)}$, $\psi_{\uparrow\downarrow+\downarrow\uparrow}^{(m_j=1)}$, and $\psi_{\downarrow\downarrow}^{(m_j=1)}$. Among these relativistically induced partial waves, the $\psi_{\uparrow\uparrow}^{(m_j=1)}$ configuration is particularly significant as it directly drives the leading-twist transverse LCDA.

Two distinct leading-twist LCDAs characterize this state, extracted from the longitudinal and transverse light-like correlators:
\begin{equation}\label{eqn:h}
 \begin{split}
     \langle0|\overline\psi(-z)\gamma^+\gamma_5\psi(+z)|h(p, \lambda=0)\rangle\big|_{z^+=z_\perp=0}
=\,& p^+f_{h} \int_0^1 \dd x \, \exp\big[ i (x-\half) p^+z^- \big] \phi_h(x), \\
     \langle0|\overline\psi(-z)\sigma^{+i}\psi(+z)|h(p, \lambda=\pm)\rangle\big|_{z^+=z_\perp=0}
=\,& p^+\epsilon^{ij}e^j_{\pm}f_{h}^\perp \int_0^1 \dd x \, \exp\big[ i (x-\half) p^+z^- \big] \phi_h^\perp(x)\;.
 \end{split}
\end{equation}
The differing quantum numbers of the interpolating currents dictate distinct normalization conditions for these LCDAs. The axial vector current $\overline\psi\gamma^+\gamma_5\psi$ has $\textsf{C} = +1$, which fundamentally mismatches the $\textsf{C} = -1$ symmetry of the $h$ meson. This forces the local matrix element to vanish $\langle0 | \overline\psi(0)\gamma^+\gamma_5\psi(0) | h\rangle = 0$, meaning $f_h$ is not a conventional decay constant, and the longitudinal LCDA must be normalized to its first moment. Conversely, the tensor current $\overline\psi\sigma^{+i}\psi$ is inherently $\textsf{C} = -1$, perfectly matching the meson. Its local matrix element is non-zero, defining a true decay constant $f_h^\perp$, allowing the transverse LCDA to be normalized to its zeroth moment:
\begin{equation}
    \begin{split}
        & \int_0^1 \dd x\, (2x-1)\phi_h(x) = 1, \\
        & \int_0^1 \dd x\, \phi_h^\perp(x) = 1.
    \end{split}
\end{equation}
Projecting these definitions onto the LFWF basis yields their integral representations:
\begin{align}
        \frac{f_h}{2\sqrt{2N_c}} \phi_h(x) =\,& \frac{1}{\sqrt{x(1-x)}} \int \frac{\dd^2k_\perp}{2(2\pi)^3} \psi_{\uparrow\downarrow-\downarrow\uparrow/h}^{(m_j=0)}(x, \vec k_\perp) \label{eqn:LCDA_h} \;,\\
        \frac{f_h^\perp}{2\sqrt{2N_c}} \phi_h^\perp(x) =\,& \frac{1}{\sqrt{x(1-x)}} \int \frac{\dd^2k_\perp}{2(2\pi)^3} \psi_{\uparrow\uparrow/h}^{(m_j=+1)}(x, \vec k_\perp). \label{eqn:LCDA_hperp} 
\end{align}

\subsection{Tensor  $T$ ($2^{++}$)}

\begin{figure}
    \centering
    \subfigure[\ $\psi_{\uparrow\downarrow+\downarrow\uparrow}^{(m_j=0)}$]{\includegraphics[width=0.3\textwidth]{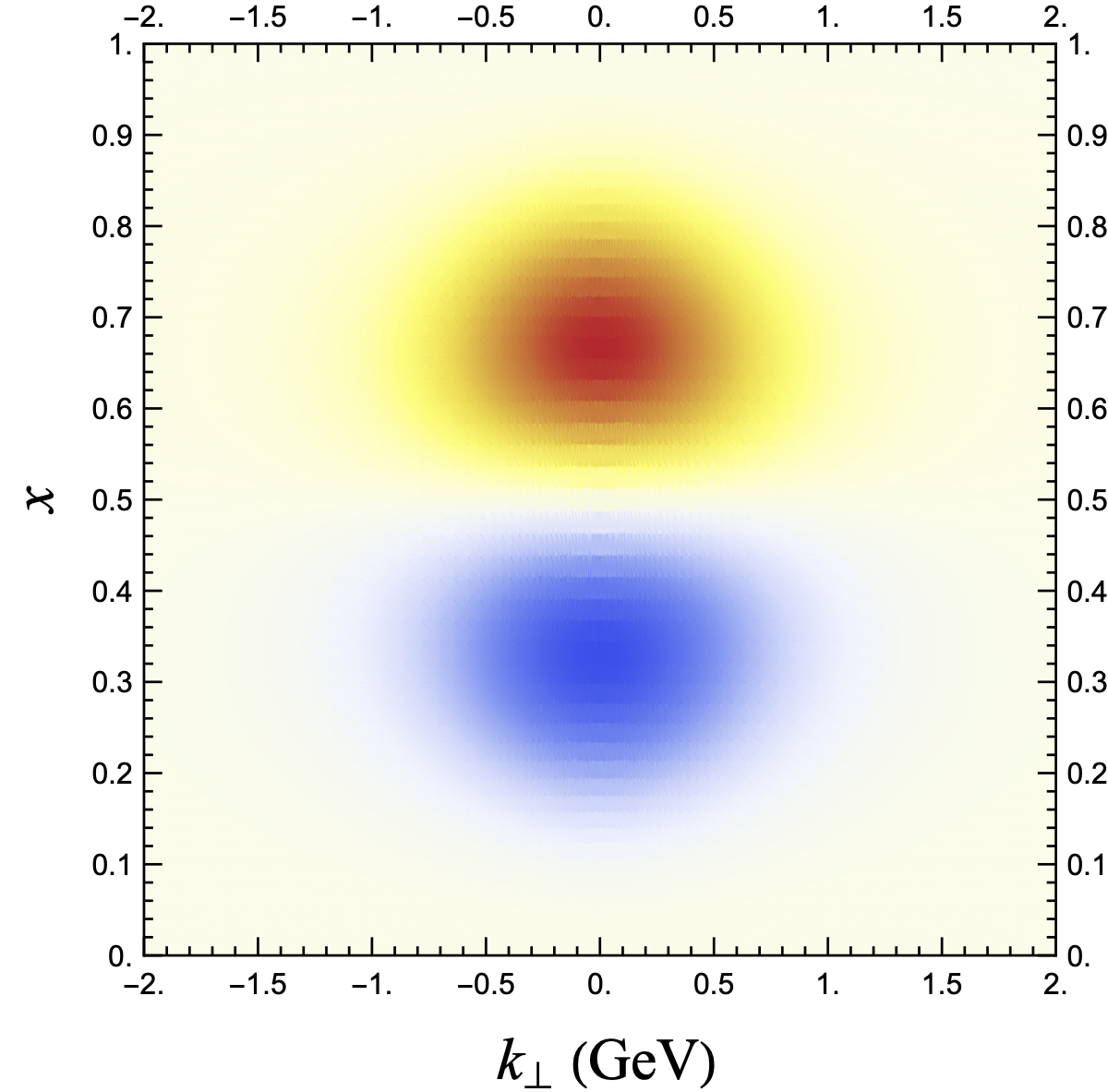}} \;
    \subfigure[\ $\psi_{\downarrow\downarrow}^{(m_j=0)}$]{\includegraphics[width=0.3\textwidth]{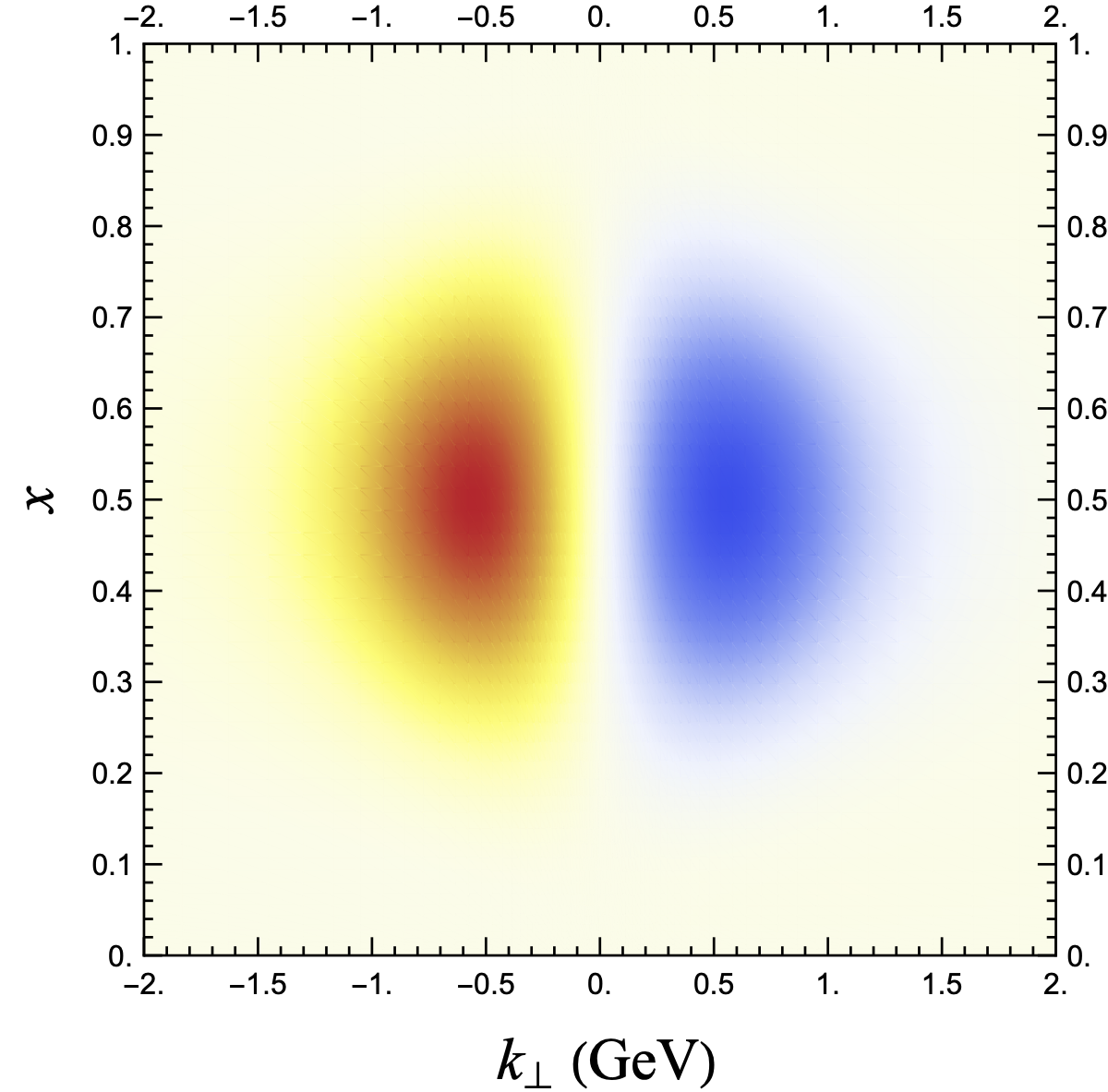}} \;
    \subfigure[\ $\psi_{\uparrow\downarrow-\downarrow\uparrow}^{(m_j=1)}$]{\includegraphics[width=0.3\textwidth]{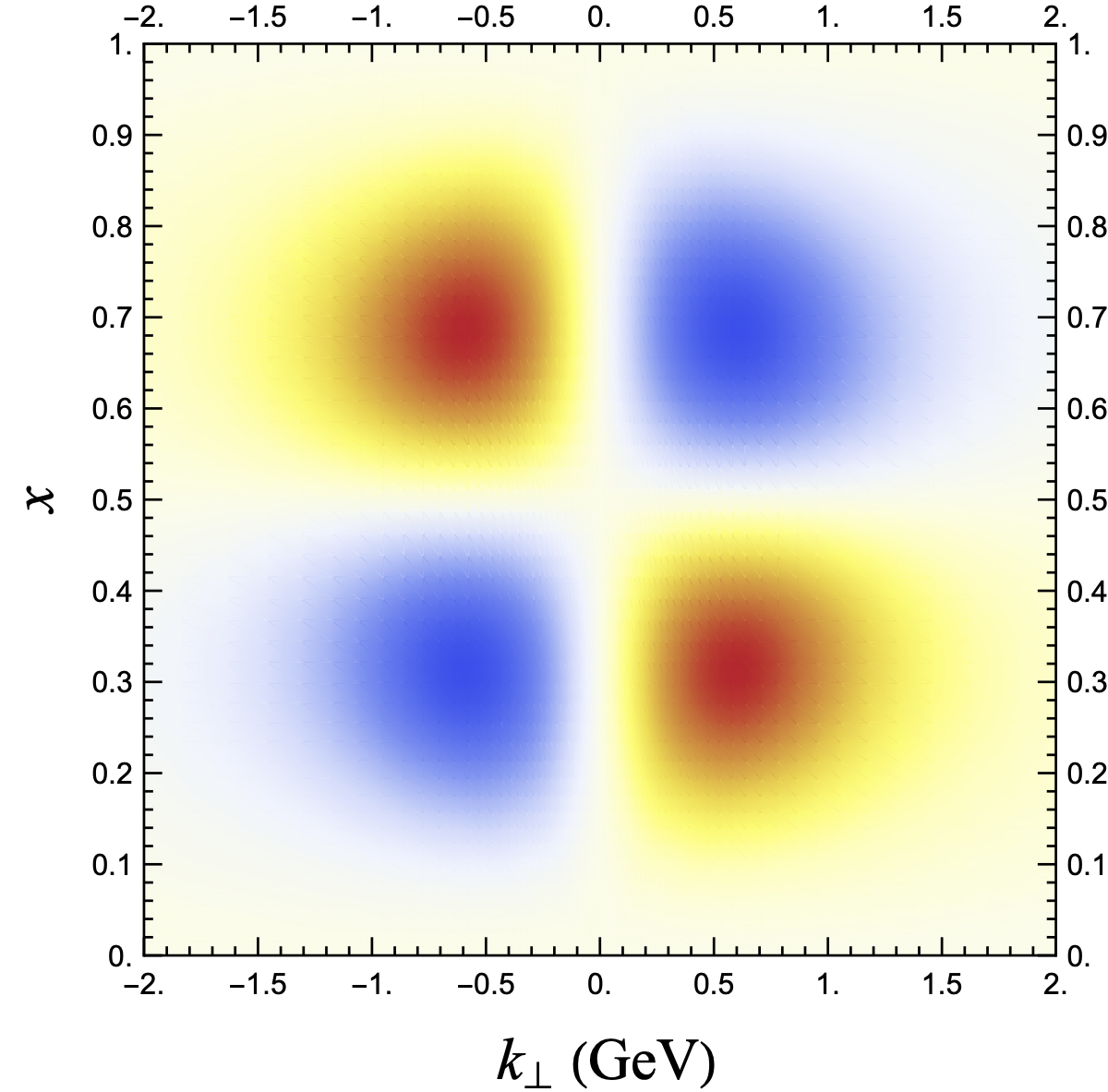}} \\
    \subfigure[\ $\psi_{\uparrow\downarrow+\downarrow\uparrow}^{(m_j=1)}$]{\includegraphics[width=0.3\textwidth]{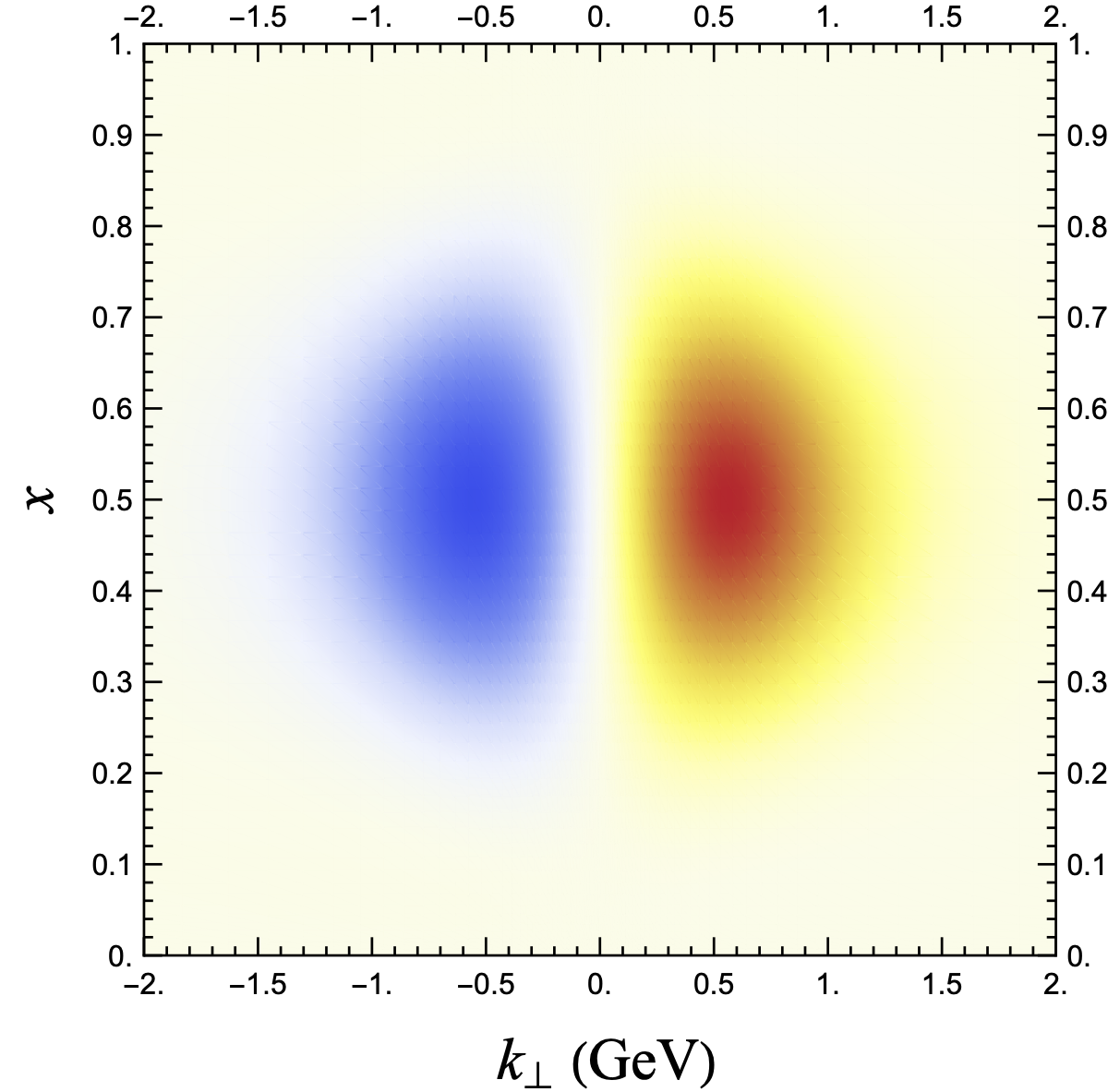}} \;
    \subfigure[\ $\psi_{\uparrow\uparrow}^{(m_j=1)}$]{\includegraphics[width=0.3\textwidth]{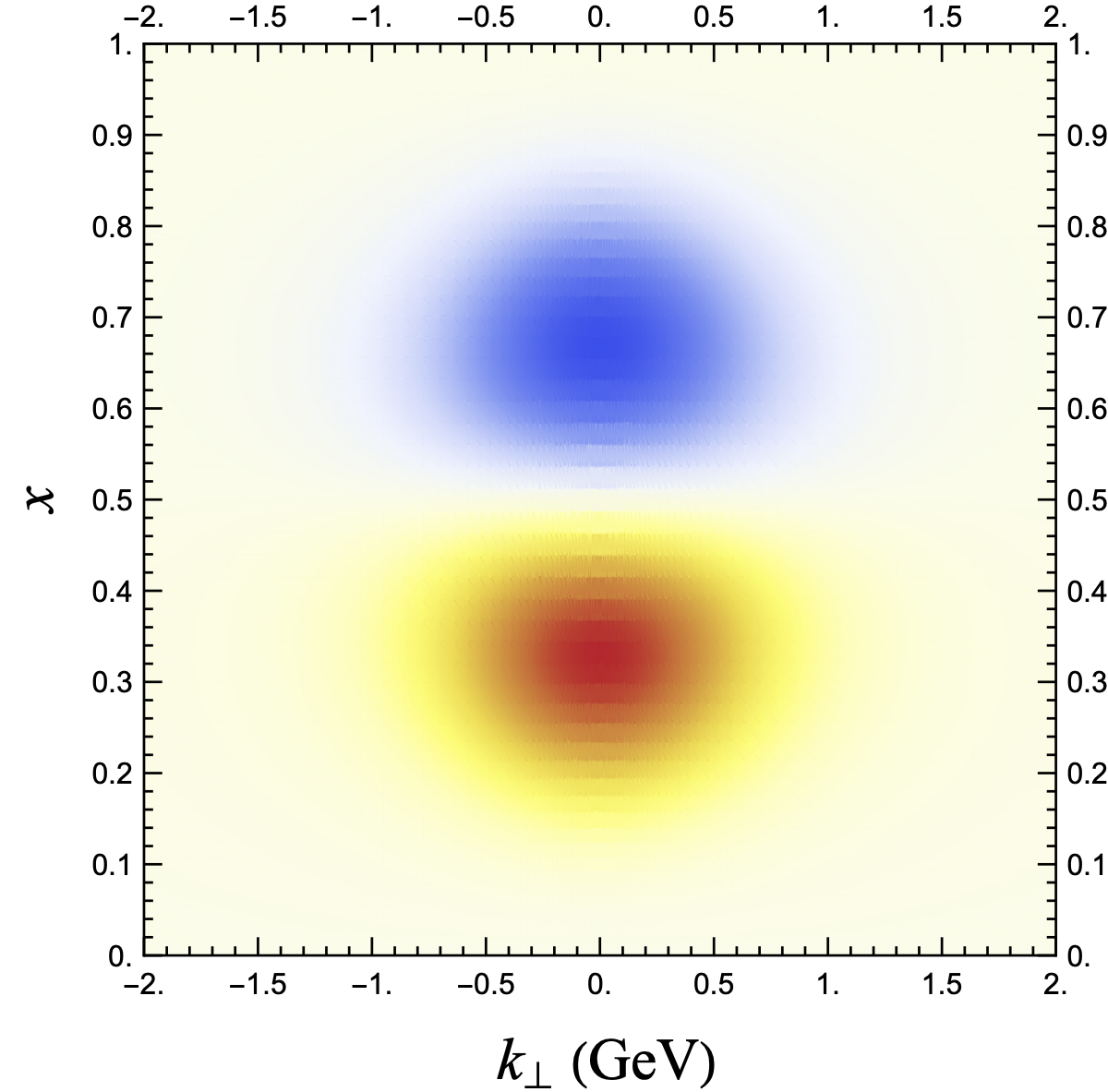}} \;
    \subfigure[\ $\psi_{\downarrow\downarrow}^{(m_j=1)}$]{\includegraphics[width=0.3\textwidth]{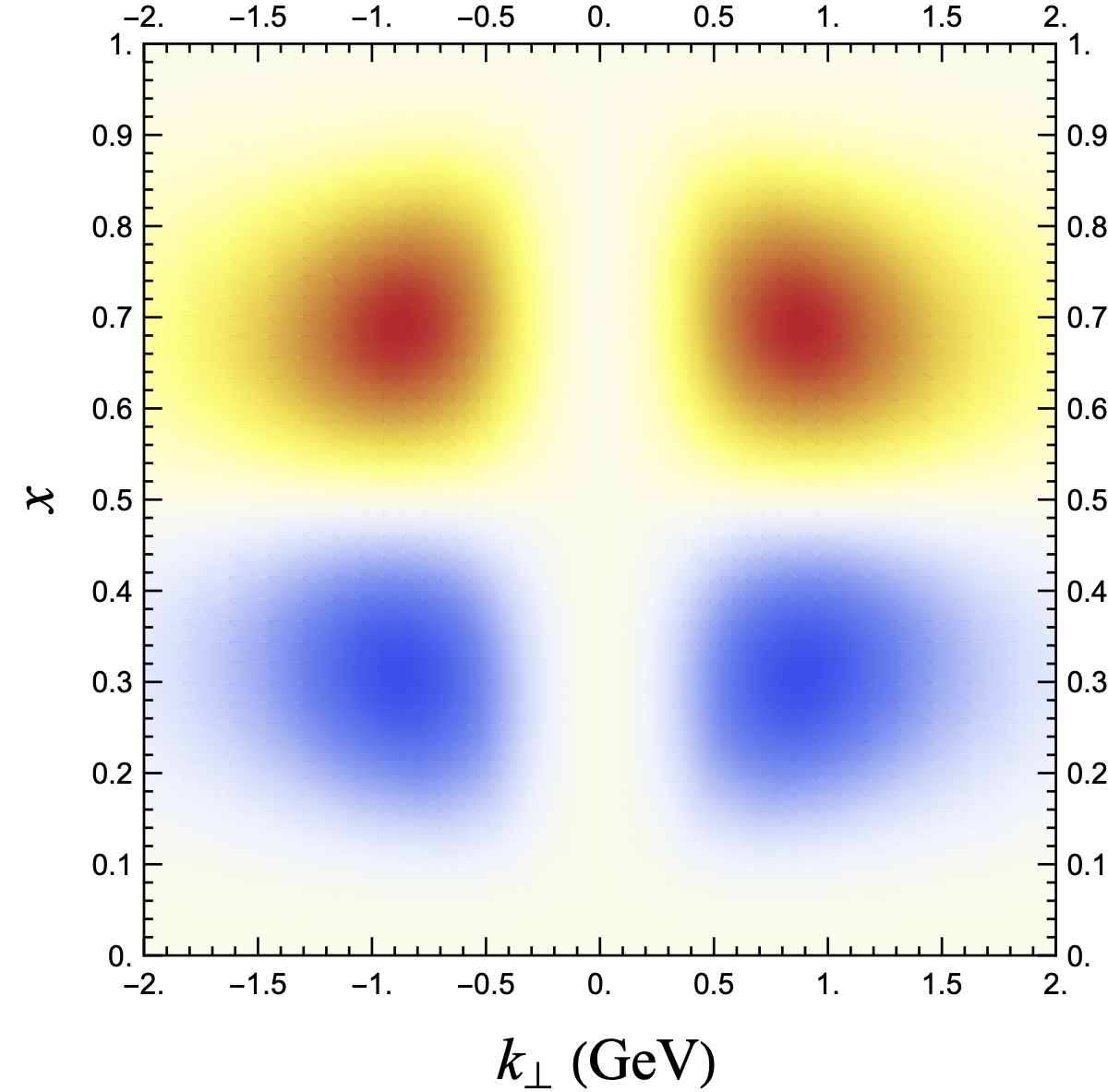}} 

    \subfigure[\ $\psi_{\uparrow\downarrow-\downarrow\uparrow}^{(m_j=2)}$]{\includegraphics[width=0.3\textwidth]{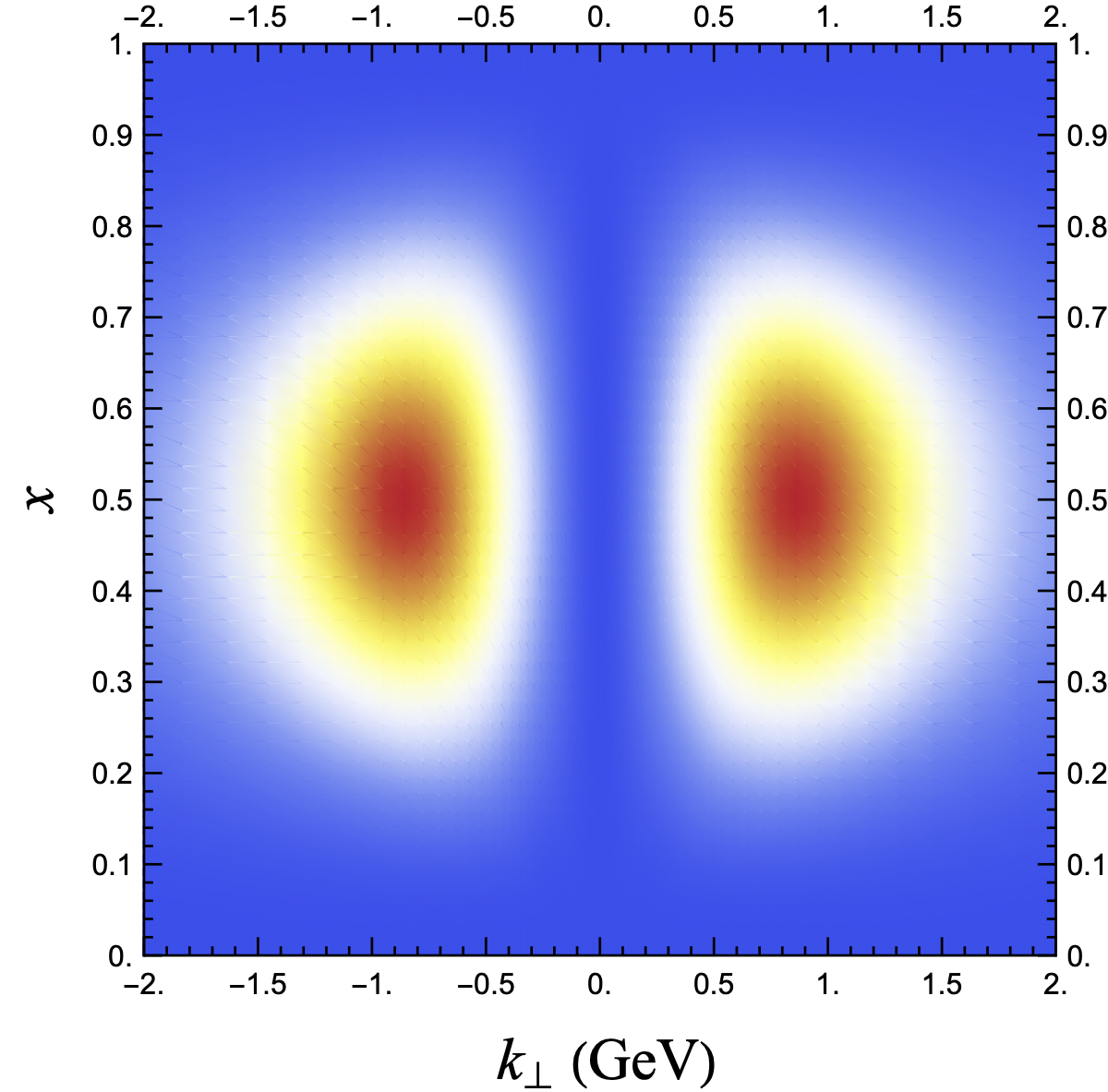}} 
    \subfigure[\ $\psi_{\uparrow\downarrow-\downarrow\uparrow}^{(m_j=2)}$]{\includegraphics[width=0.3\textwidth]{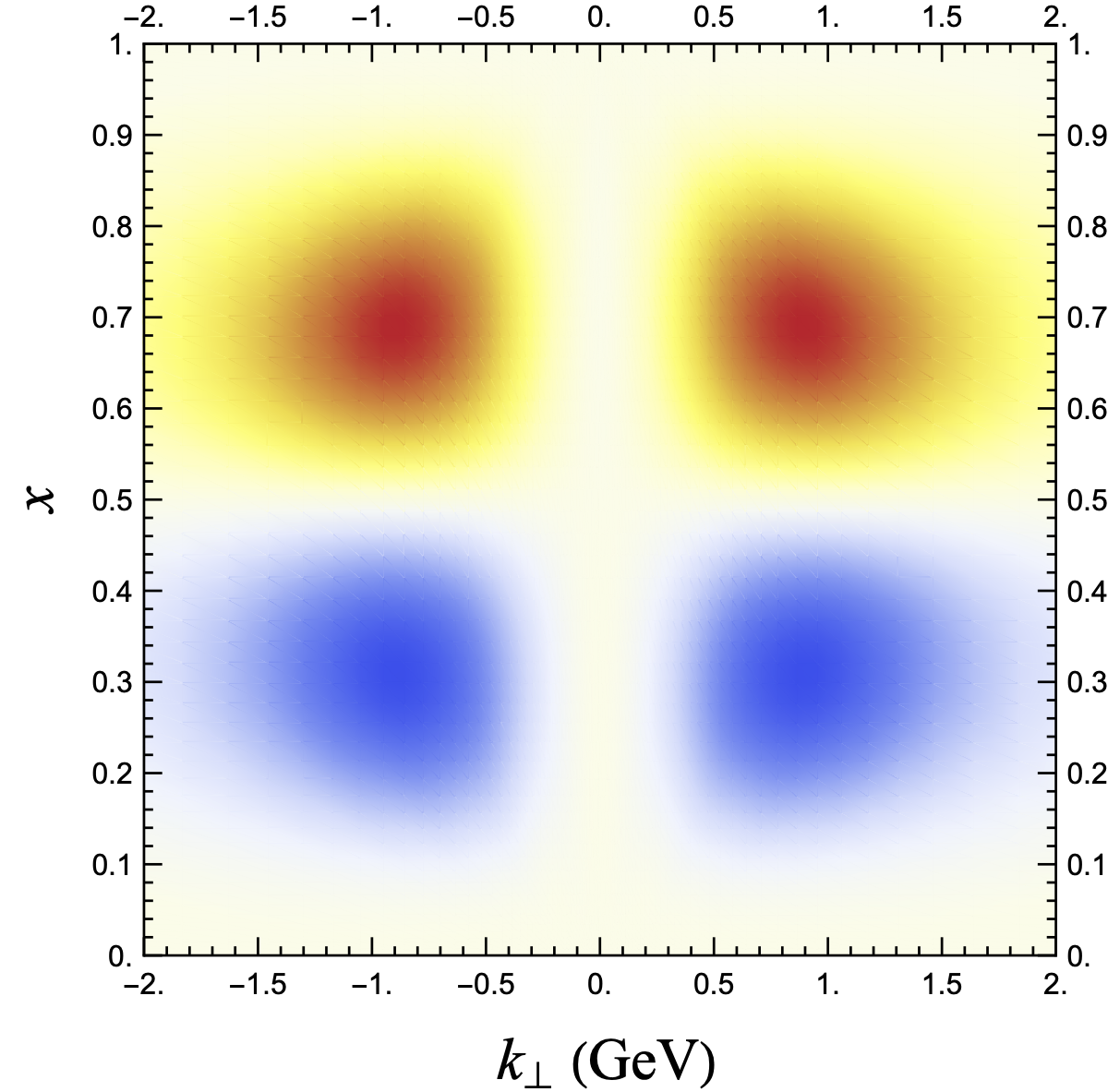}} 
    \subfigure[\ $\psi_{\uparrow\uparrow}^{(m_j=2)}$]{\includegraphics[width=0.3\textwidth]{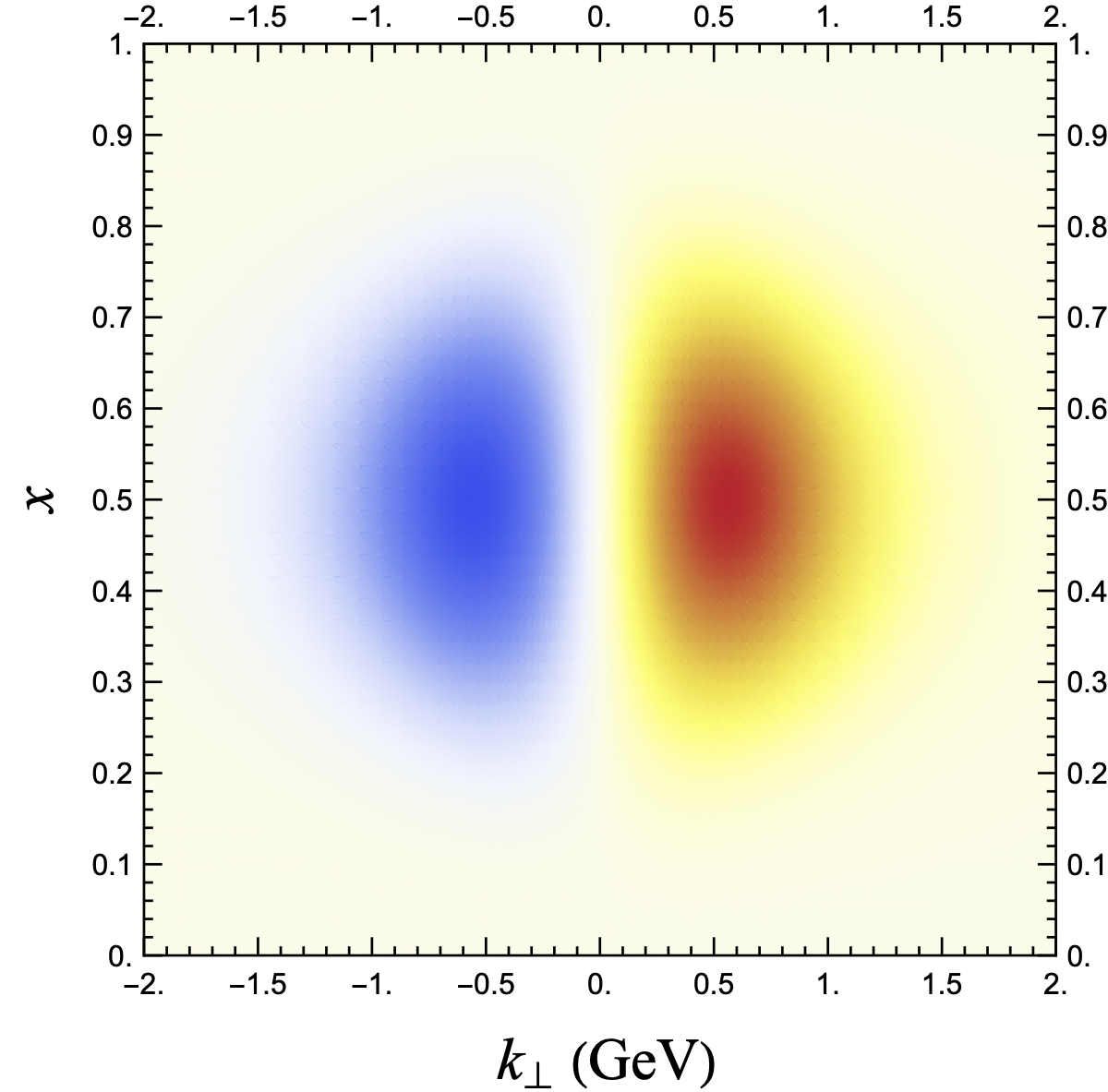}} 
    \subfigure[\ $\psi_{\downarrow\downarrow}^{(m_j=2)}$]{\includegraphics[width=0.3\textwidth]{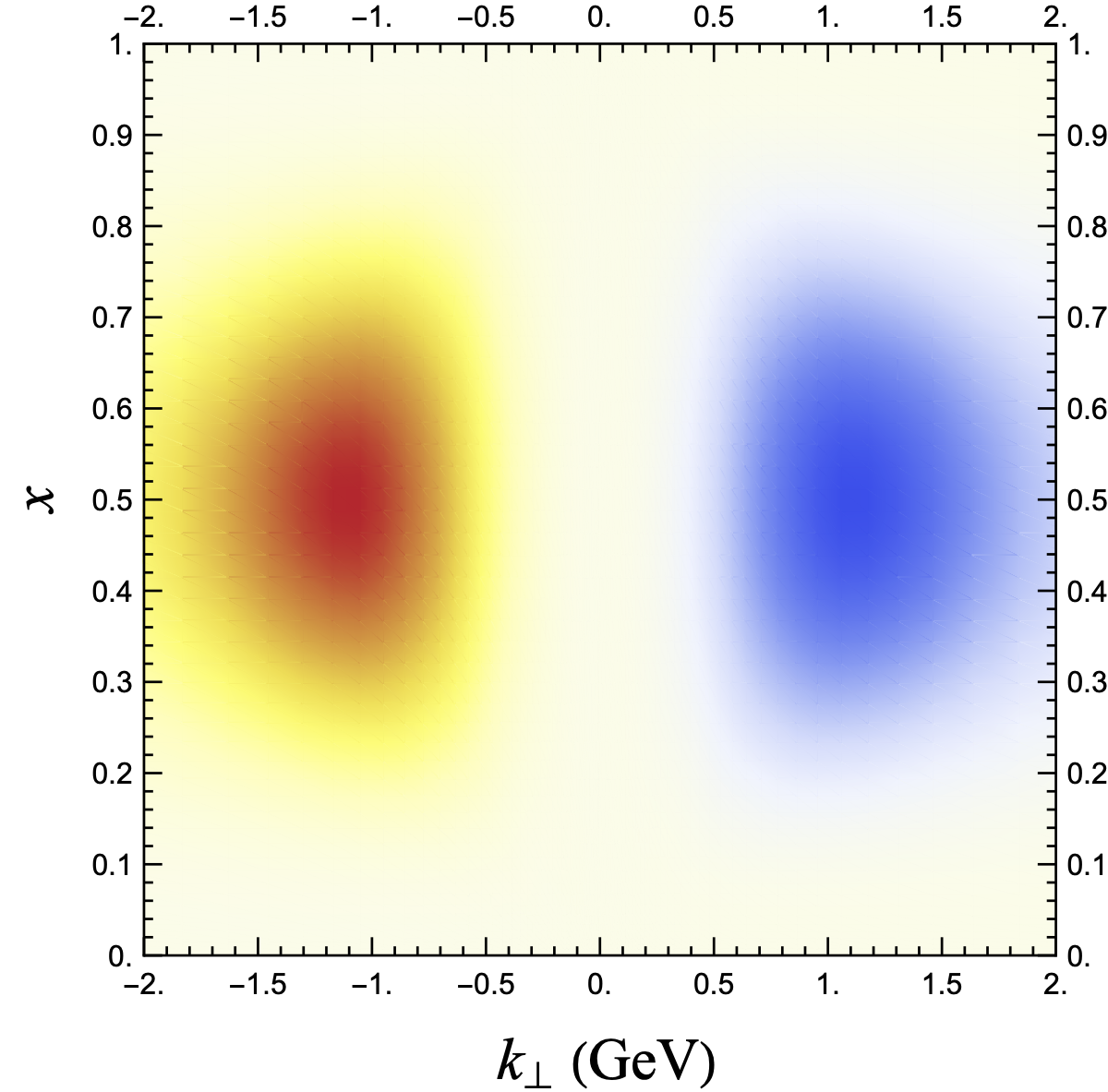}} 

    \caption{Light-front wave functions of  tensor quarkonium $\chi_{c2}(1P)$ obtained from BLFQ. }
    \label{fig:chic2}
\end{figure}

In the non-relativistic constituent quark model, the $2^{++}$ tensor meson is a purely ${}^3P_2$ state ($L=1, S=1$). Its wave function takes the form:
\begin{equation}
\Psi_{s\bar s/T}^\lambda(\vec k) = \sqrt{\frac{3}{8\pi}} \big[\hat k \cdot \tensor\xi^\lambda\cdot \vec\sigma \sigma_2\big]_{s\bar s}\phi_T(|\vec k|)\;,
\end{equation}
where $\tensor \xi^\lambda$ (with $\lambda=0,\pm 1, \pm 2$) represents a set of symmetric and traceless unit polarization tensors satisfying the completeness relation:
\begin{equation}
  \sum_{\lambda=0,\pm 1, \pm 2} \xi_{ij}^{\lambda*}\xi_{nm}^{\lambda} = \frac{1}{2}(\delta_{im}\delta_{jn} + \delta_{in}\delta_{jm}) - \frac{1}{3}\delta_{ij}\delta_{nm}.
\end{equation}
On the light front, the general covariant structure of a tensor meson $T$ with momentum $P$, mass $M_T$, and polarization $\lambda$ is governed by ten invariant functions $\phi_{1-10}$:
\begin{eqnarray}
\psi_{s\overline{s}/T}^{\lambda}(x,\vec{k}_{\perp}) &=& e_{\mu\nu}^{\lambda}(P)\overline{u}_{s}(p)\bigg[ \gamma^{\mu}(p-\overline{p})^{\nu}\phi_{1}(x,k_{\perp}) + \gamma^{\mu}\frac{\omega^{\nu}}{P^{+}}\phi_{2}(x,k_{\perp}) \nonumber \\
&&+ (p-\overline{p})^{\mu}(p-\overline{p})^{\nu}\phi_{3}(x,k_{\perp}) + (p-\overline{p})^{\mu}\frac{\omega^{\nu}}{P^{+}}\phi_{4}(x,k_{\perp}) + \frac{\omega^{\mu}\omega^{\nu}}{(P^{+})^{2}}\phi_{5}(x,k_{\perp}) \nonumber \\
&&+ \frac{\gamma^{+}}{P^{+}}\gamma^{\mu}(p-\overline{p})^{\nu}\phi_{6}(x,k_{\perp}) + \frac{\gamma^{+}}{(P^{+})^{2}}\gamma^{\mu}\omega^{\nu}\phi_{7}(x,k_{\perp}) \nonumber \\
&&+ \frac{\gamma^{+}}{P^{+}}(p-\overline{p})^{\mu}(p-\overline{p})^{\nu}\phi_{8}(x,k_{\perp}) + \frac{\gamma^{+}}{(P^{+})^{2}}(p-\overline{p})^{\mu}\omega^{\nu}\phi_{9}(x,k_{\perp}) \nonumber \\
&&+ \frac{\gamma^{+}}{(P^{+})^{3}}\omega^{\mu}\omega^{\nu}\phi_{10}(x,k_{\perp}) \bigg] v_{\overline{s}}(\overline{p})\;,
\end{eqnarray}
where $e_{\mu\nu}^{\lambda}(P)$ is the symmetric, traceless polarization tensor, and the kinematic dependence of $\phi_i(x, k_\perp)$ is implicitly understood.

Two leading-twist LCDAs characterize the tensor meson, extracted from the longitudinal vector and transverse tensor correlators:
\begin{align}
\langle 0 | \bar{\psi}(-z) \gamma^+ \psi(+z) | T(P, \lambda) \rangle \big|_{z^+ = z_\perp = 0} =\,& f_T M_T^2 \frac{e^{++}(\lambda)}{P^+} \int_0^1 \dd x \, e^{i (x - 1/2) P^+ z^-} \phi_T(x), \\
\langle 0 | \bar{\psi}(-z) \sigma^{+\perp} \psi(+z) | T(P, \lambda) \rangle \big|_{z^+ = z_\perp = 0} =\,& i f_T^\perp M_T e^{+\perp}(\lambda) \int_0^1 dx \, e^{i (x - 1/2) P^+ z^-} \phi_T^\perp(x)\;.
\end{align}
Because the $2^{++}$ tensor meson is strictly $\mathsf{C}$-even ($+1$), and both the vector current $\bar{\psi}\gamma^\mu\psi$ and the tensor current $\bar{\psi}\sigma^{\mu\nu}\psi$ are inherently $\mathsf{C}$-odd ($-1$), their local matrix elements ($z \to 0$) must identically vanish $\langle 0 | \mathcal{J} | T \rangle = 0$. Consequently, neither correlator yields a conventional zeroth-moment decay constant. Instead, both the longitudinal and transverse LCDAs are strictly antisymmetric under longitudinal momentum fraction exchange $\phi(1-x) = -\phi(x)$. Therefore, both LCDAs must be normalized by their first moments:
\begin{align}
\int_0^1 \dd x \, (2x - 1) \phi_T(x) =&\, 1, \\
\int_0^1 \dd x \, (2x - 1) \phi_T^\perp(x) =&\, 1.
\end{align}

Projecting these constraints onto the LFWF basis yields their integral representations:
\begin{align}
\frac{f_T}{2\sqrt{2N_c}} \phi_T(x) =\,& \frac{1}{\sqrt{x(1-x)}} \int\frac{\dd^2 {k}_\perp}{2(2\pi)^3} \psi_{\uparrow\downarrow+\downarrow\uparrow/T}^{\lambda=0}(x, \vec{k}_\perp), \label{eqn:LCDA_T}\\ 
\frac{f_T^\perp}{2\sqrt{2N_c}} \phi_T^\perp(x) =\,& \frac{1}{\sqrt{x(1-x)}} \int\frac{\dd^2 {k}_\perp}{2(2\pi)^3} \psi_{\uparrow\uparrow/T}^{\lambda=+1}(x, \vec{k}_\perp). \label{eqn:LCDA_Tperp}
\end{align}

\section{Numerical results} \label{sec:numerical_results}

The LFWFs in this study are generated using a phenomenological Hamiltonian inspired by light-front holographic QCD \cite{Li:2017mlw}. The effective Hamiltonian incorporates a confining interaction and a one-gluon exchange (OGE) term. The confinement strength $\kappa$ and the quark mass $m_q$ are fixed by fitting the quarkonium mass spectrum, achieving an rms deviation of approximately 40 MeV from the Particle Data Group (PDG) values for states below the open-flavor thresholds.
These LFWFs are obtained as the eigenfunctions of the light-front invariant mass squared operator, $H_\textsc{lc} \equiv P^+ P^- - \vec P^2_\perp$:
\begin{equation}\label{eqn:LC_Schrodinger}
H_\textsc{lc} |\psi_h(p, j, m_j)\rangle = M^2_h |\psi_h(p, j, m_j)\rangle,
\end{equation}
where $M_h$, $j$, and $m_j$ are the hadron's invariant mass, total spin, and its magnetic spin projection, respectively. The kinematic momenta are defined as $P^\pm \equiv P^0 \pm P^3$ and $\vec P_\perp = (P^1, P^2)$, with $P^-$ serving as the light-front Hamiltonian generating evolution along the light-front time $x^+ \equiv x^0 + x^3$ \cite{Li:2017mlw}.

Restricting the Fock space to the valence ($q\bar{q}$) sector, a sufficient and standard approximation for heavy quarkonia, the meson state vector expands as:
\begin{equation}
\begin{split}
|\psi_h(p, j, m_j)\rangle 
 =\,& \sum_{s, \bar s} \int_0^1\frac{\dd x}{2x(1-x)} \int\frac{\dd^2k_\perp}{(2\pi)^3} \psi^{(j,m_j)}_{s\bar s/h}(x, \vec k_\perp) \\
 & \times \frac{1}{\sqrt{N_c}} \sum_i b^\dagger_{s i}\big(xp^+, \vec k_\perp+x\vec p_\perp\big) d^\dagger_{\bar s i}\big((1-x)p^+, -\vec k_\perp+(1-x)\vec p_\perp\big) |0\rangle\;.
\end{split}
\end{equation}
Here, $\psi^{(j,m_j)}_{s\bar s/h}(x, \vec k_\perp)$ represents the valence LFWF, while $b^\dagger$ and $d^\dagger$ denote the quark and antiquark creation operators. This definition imposes the standard LFWF unit normalization:
\begin{equation}
\sum_{s, \bar s} \int_0^1\frac{\dd x}{2x(1-x)}\int\frac{\dd^2 k_\perp}{(2\pi)^3}  
\big| \psi^{(j,m_j)}_{s\bar s/h}(x, \vec k_\perp) \big|^2
= 1.
\end{equation}

The discrete symmetries governing these states on the light front are mirror parity, $m_\textsf{P} = (-i)^{2j}\textsf{P}$, and charge conjugation, $\textsf{C}$. Their LFWF representations evaluate to:
\begin{align}
    m_{\textsf{P}} =\,& \sum_{s, \bar s} \int \frac{\dd x}{2x(1-x)} \int \frac{\dd^2 k_\perp}{(2\pi)^3}  {\psi}_{-s-\bar s}^{(j,m_j)*}(x, \tilde k_\perp) \psi_{s\bar s}^{(j,-m_j)}(x, \vec k_\perp), \\
    \textsf{C} =\,& -\sum_{s, \bar s} \int \frac{\dd x}{2x(1-x)} \int \frac{\dd^2k_\perp}{(2\pi)^3}  {\psi}_{\bar ss}^{(j,m_j)*}(1-x, -\vec k_\perp) \psi_{s\bar s}^{(j,m_j)}(x, \vec k_\perp),
\end{align}
where $\tilde k_\perp = (-k_x, k_y)$. Because the mirror parity operation actively flips the magnetic spin projection ($m_j \to -m_j$), this symmetry constraint is generally enforced only within the $m_j=0$ sector.

Equation (\ref{eqn:LC_Schrodinger}) is solved numerically using the BLFQ framework. The wave functions are mapped onto a chosen basis space:
\begin{equation}
    \psi^{(j,m_j)}_{s\bar s/h}(x, \vec k_\perp) = \sum_{n,m,l} \psi^{(j,m_j)}_{s\bar s/h}(n,m,l) \phi_{nm}(\vec k_\perp/\sqrt{x(1-x)}) \chi_l(x)\;,
\end{equation}
where $\vec q_\perp = \vec k_\perp/\sqrt{x(1-x)}$ is the invariant transverse momentum variable. The transverse ($\phi_{nm}$) and longitudinal ($\chi_l$) basis functions are the analytical solutions of the effective Hamiltonian in the absence of the OGE term:
\begin{align}
     & \phi_{nm}(\vec k_\perp)=\kappa^{-1}\sqrt{\frac{4\pi n!}{(n+|m|)!}}{\bigg(\frac{k_\perp}{\kappa}\bigg)}^{|m|}
  \exp(-k_\perp^2/(2\kappa^2))
  L_n^{|m|}(k_\perp^2/\kappa^2)\exp(i m\theta_k), \\
  & \chi_l(x)=\sqrt{4\pi(2l+\alpha+\beta+1)}\sqrt{\frac{\Gamma(l+1)\Gamma(l+\alpha+\beta+1)}{\Gamma(l+\alpha+1)\Gamma(l+\beta+1)}}
  x^{\beta/2}{(1-x)}^{\alpha/2}P_l^{(\alpha,\beta)}(2x-1)\;,
\end{align}
where $q_\perp= |\vec q_\perp | $ and $\theta_q=\arg \vec q_\perp$. The functions $L_n^{|m|}(z)$ and $P_l^{(\alpha,\beta)}(z)$ are the associated Laguerre and Jacobi polynomials, respectively, while $\kappa$, $\alpha$, and $\beta$ are the basis scale parameters detailed in Ref.~\cite{Li:2017mlw}. The basis functions are shown to be a first approximation to QCD \cite{Li:2021jqb, deTeramond:2021yyi, Li:2021cwv, Forshaw:2024mrh, Freese:2026ulx, Hemmati:2026fqj}. 

The basis coefficients $\psi^{(j,m_j)}_{s\bar s/h}(n,m,l)$ are obtained by diagonalizing the full effective Hamiltonian matrix within a finite basis space. This space is bounded by the truncations $\sum_i (2n_i+\vert{}m_i\vert{}+1) \le N_{\max}$ (for transverse modes) and $l\le L_{\max}$ (for longitudinal modes). The exact values of these extracted coefficients are available via Mendeley Data \cite{Data:Li_2019}. 
The transverse truncation $N_\text{max}$ naturally imposes an ultraviolet (UV) cutoff on the momentum space resolution, $\Lambda_\text{UV}=\kappa\sqrt{N_\text{max}}$. Guided by previous heavy quarkonium studies \cite{Li:2017mlw, Tang:2018myz, Tang:2020org, Li:2021ejv, Wang:2023nhb}, we select basis sizes that yield a UV cutoff roughly equivalent to the mass of the respective quarkonium state ($\Lambda_\text{UV} \approx M_{Q\bar{Q}}$). For charmonium, we use $N_\text{max}=8$, yielding a resolution of $\Lambda_\text{UV} = 2.8 \,\mathrm{GeV}$. For bottomonium, we require a larger basis of $N_\text{max}=32$, which provides a corresponding resolution of $\Lambda_\text{UV} = 7.8 \,\mathrm{GeV}$.

Building upon the preceding derivations (\ref{eqn:LCDA_S}, \ref{eqn:LCDA_A}, \ref{eqn:LCDA_Aperp}, \ref{eqn:LCDA_h}, \ref{eqn:LCDA_hperp}, \ref{eqn:LCDA_T}, \ref{eqn:LCDA_Tperp}), the leading-twist LCDAs for any $P$-wave quarkonium state $M$ can be universally expressed as the transverse momentum integral of its dominant $L_z = 0$ LFWF:
\begin{equation}
        \frac{f_M}{2\sqrt{2N_c}} \phi_M(x) = \frac{1}{\sqrt{x(1-x)}} \int \frac{\dd^2k_\perp}{2(2\pi)^3} \psi_{M}^{(L_z = 0)}(x, \vec k_\perp). 
\end{equation}

\begin{figure}
    \centering
    \includegraphics[width=0.45\textwidth]{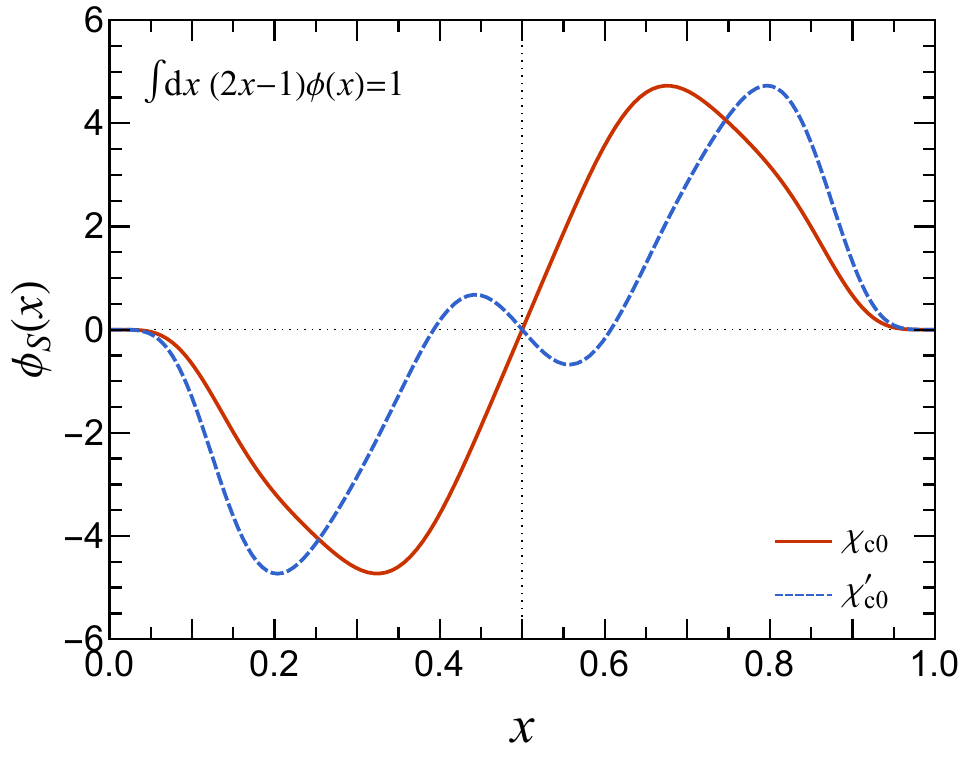}
    \includegraphics[width=0.46\textwidth]{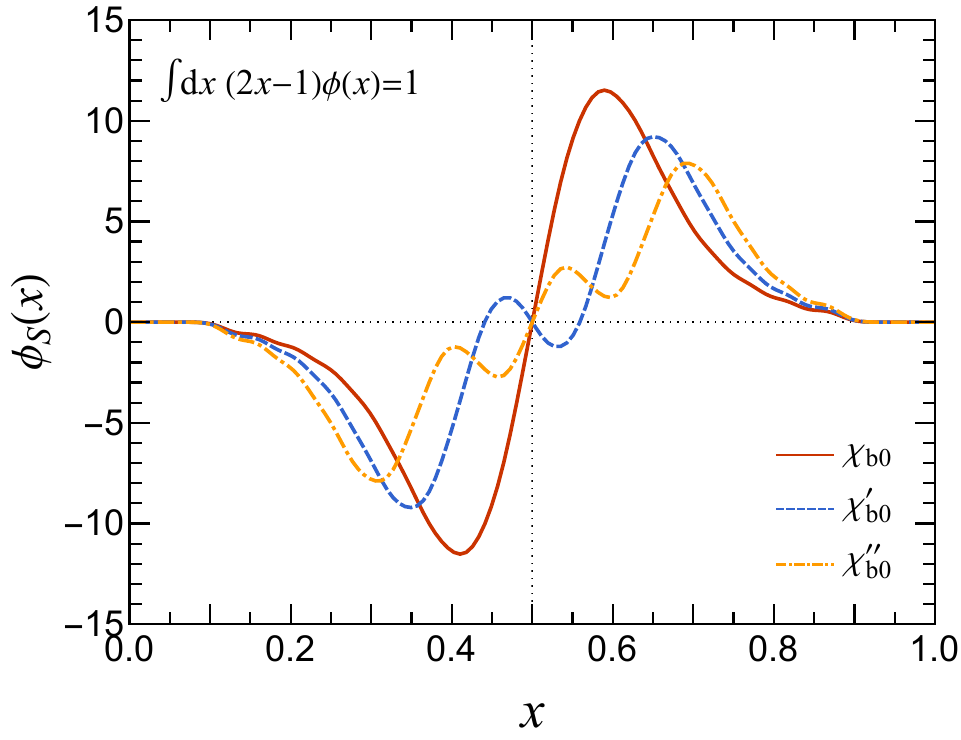}
    \caption{LCDAs of the $P$-wave scalar quarkonia $\chi_{c0}, \chi_{b0}$ and their radial excitations $\chi'_{c0}, \chi'_{b0}, \chi''_{b0}$.}
    \label{fig:S_DAs}
\end{figure}

Figure~\ref{fig:S_DAs} displays the LCDAs for the $P$-wave scalar quarkonia ($\chi_{c0}$ and $\chi_{b0}$), encompassing both ground states and radial excitations. As expected, radially excited states exhibit additional nodes governed by their radial quantum numbers.

\begin{figure}
    \centering
    \includegraphics[width=0.48\textwidth]{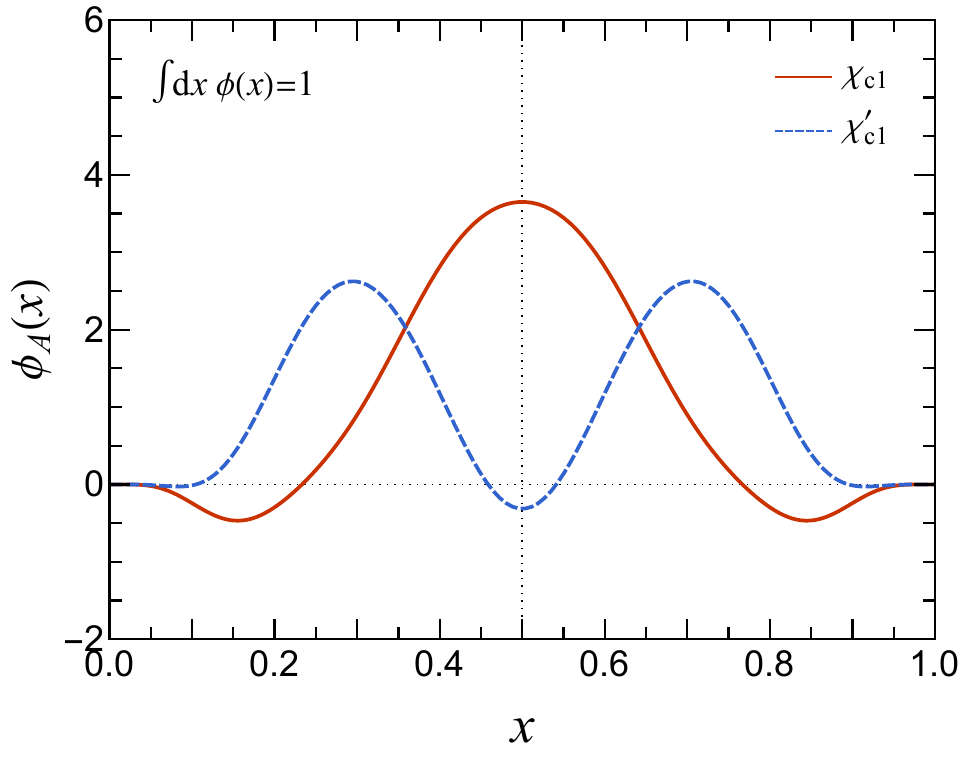}
    \includegraphics[width=0.48\textwidth]{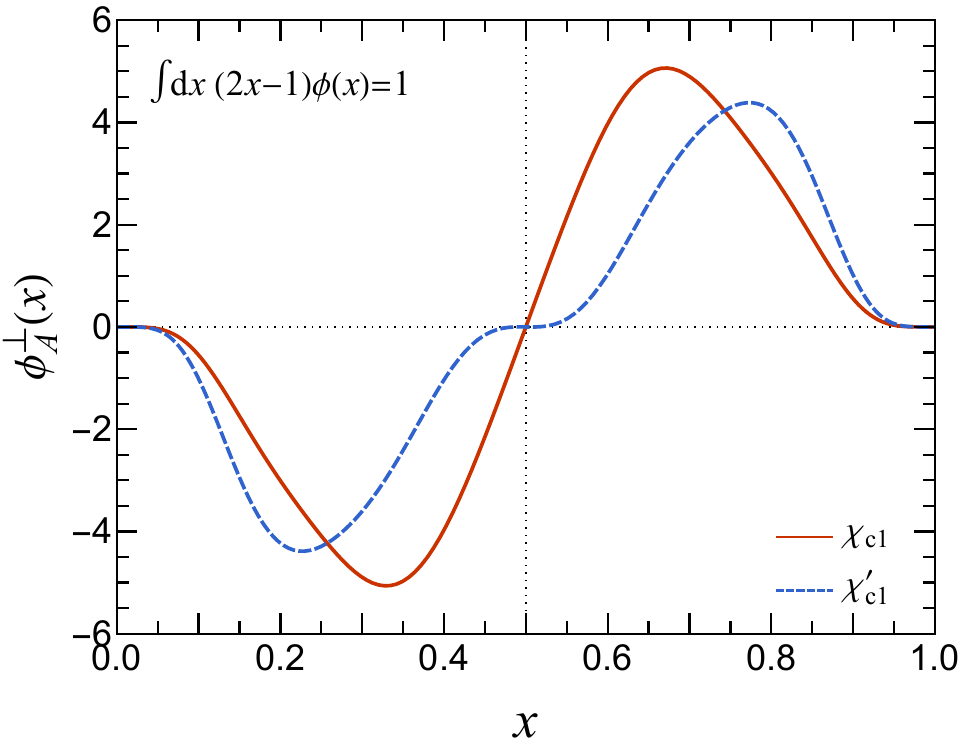} \\
   \includegraphics[width=0.48\textwidth]{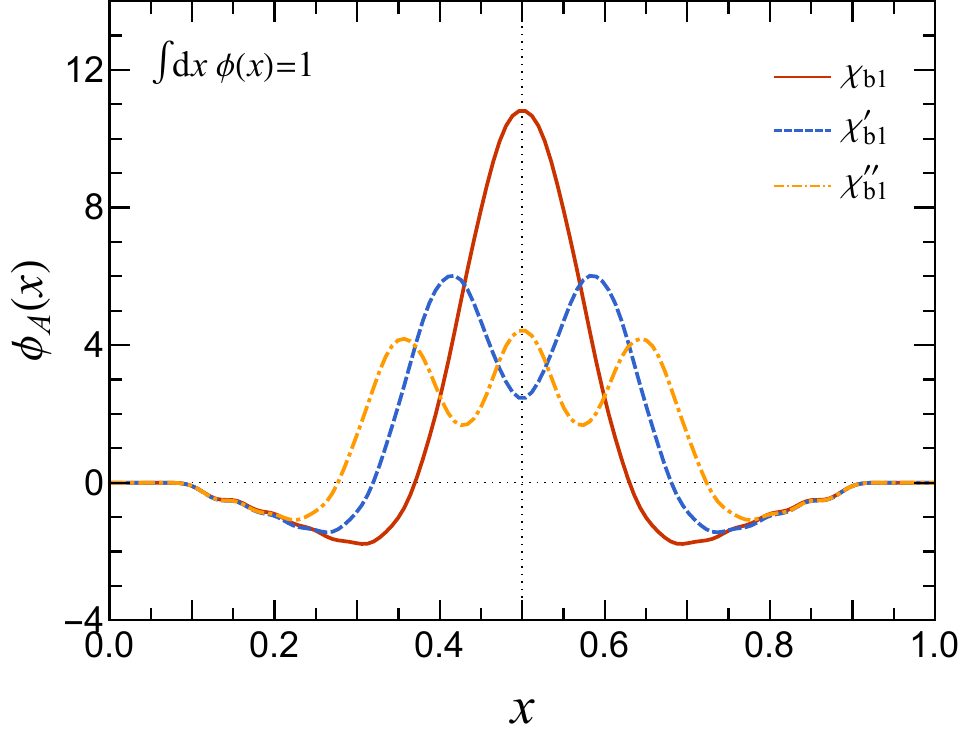}
    \includegraphics[width=0.49\textwidth]{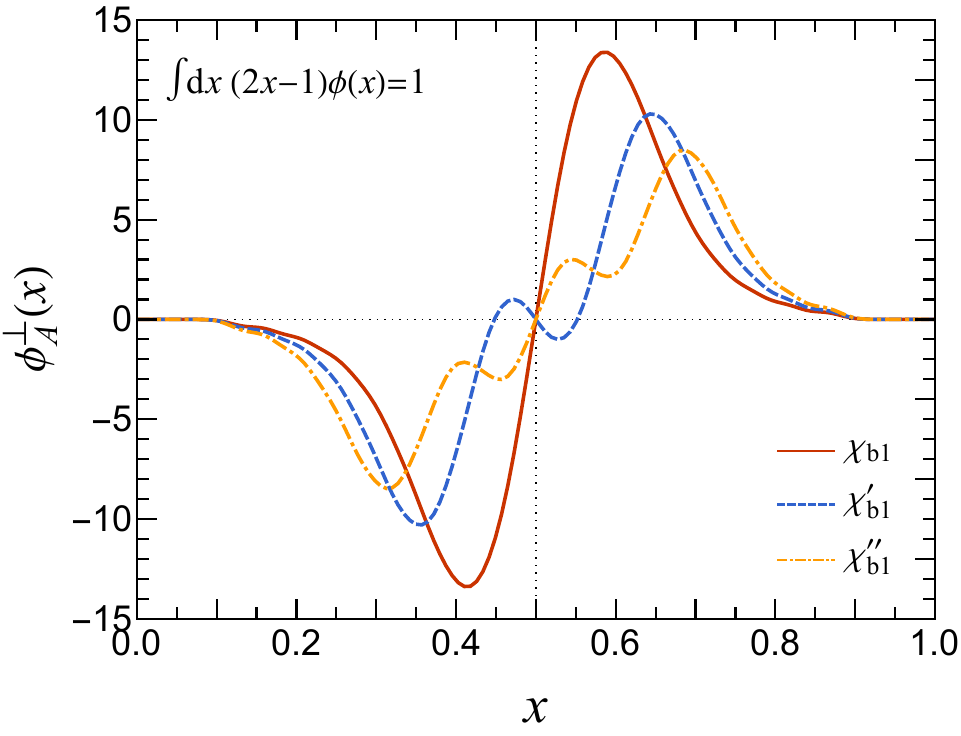}    
    \caption{Leading-twist distribution amplitudes of the $P$-wave axial vector quarkonia ($1^{++}$) $\chi_{c1}, \chi_{b1}$ and their radial excitations.}
    \label{fig:A_DAs}
\end{figure}

\begin{figure}
    \centering
    \includegraphics[width=0.48\textwidth]{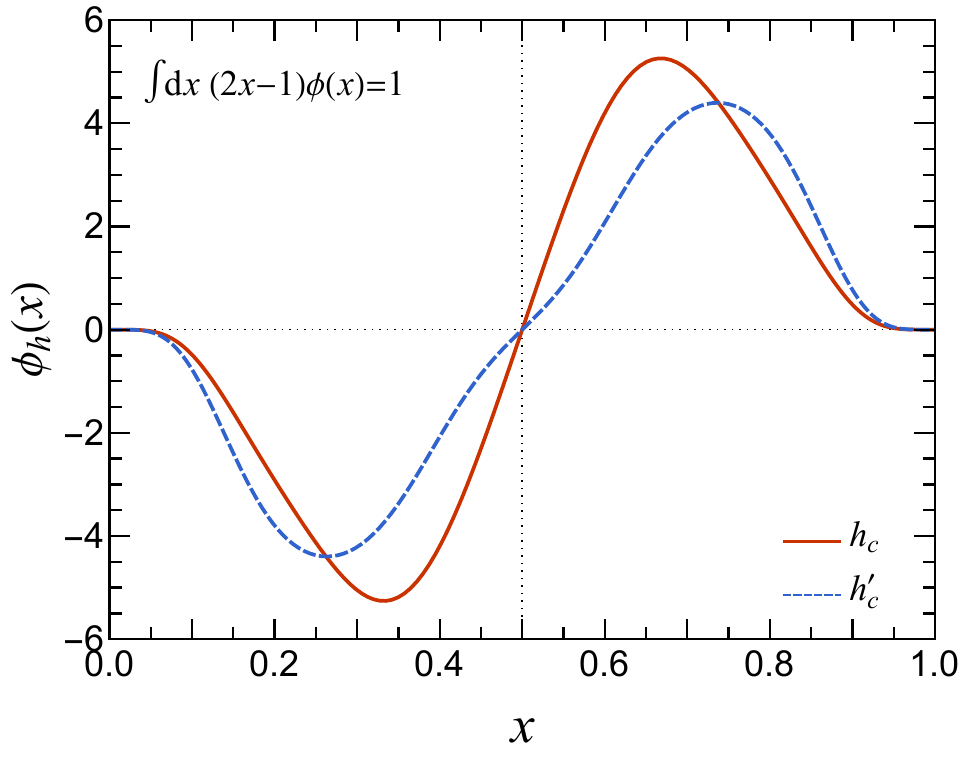}
    \includegraphics[width=0.48\textwidth]{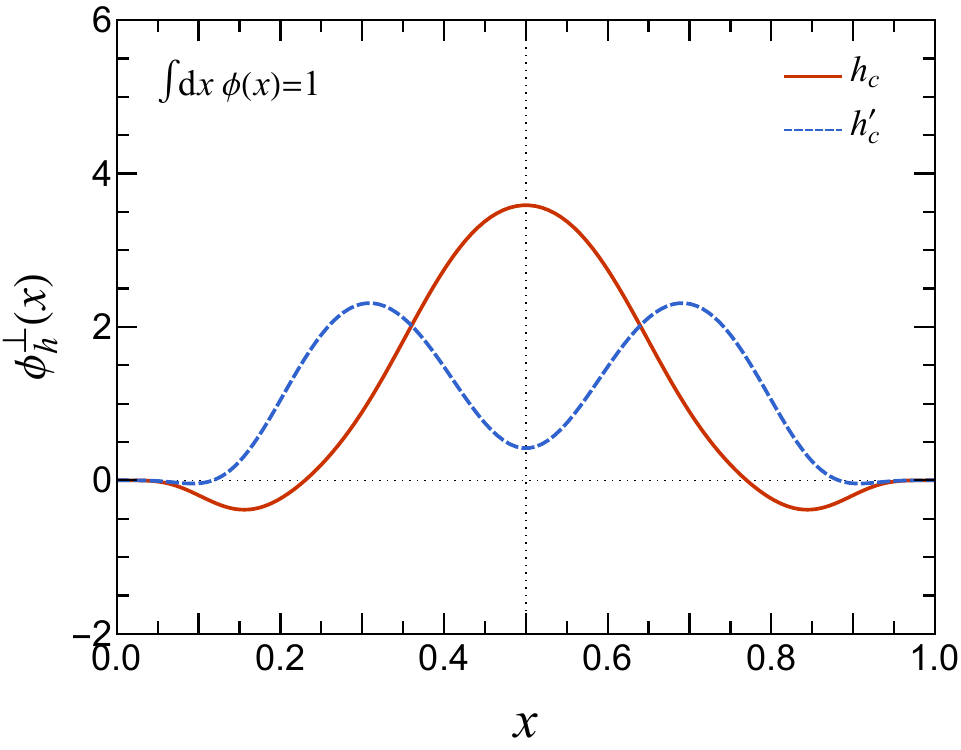} \\
   \includegraphics[width=0.48\textwidth]{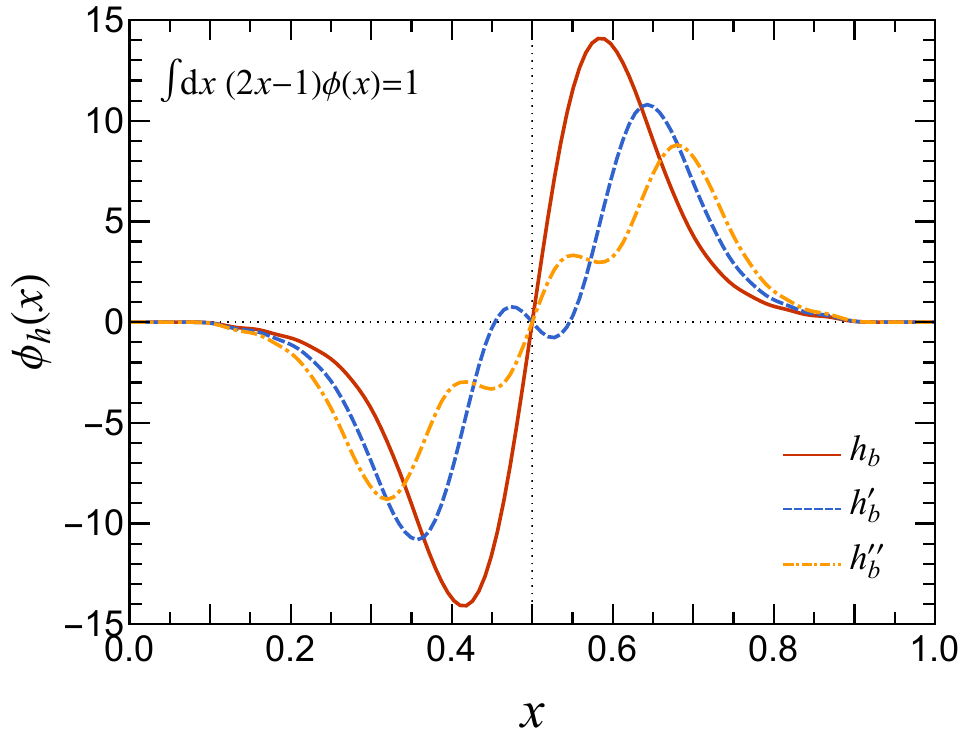}
    \includegraphics[width=0.48\textwidth]{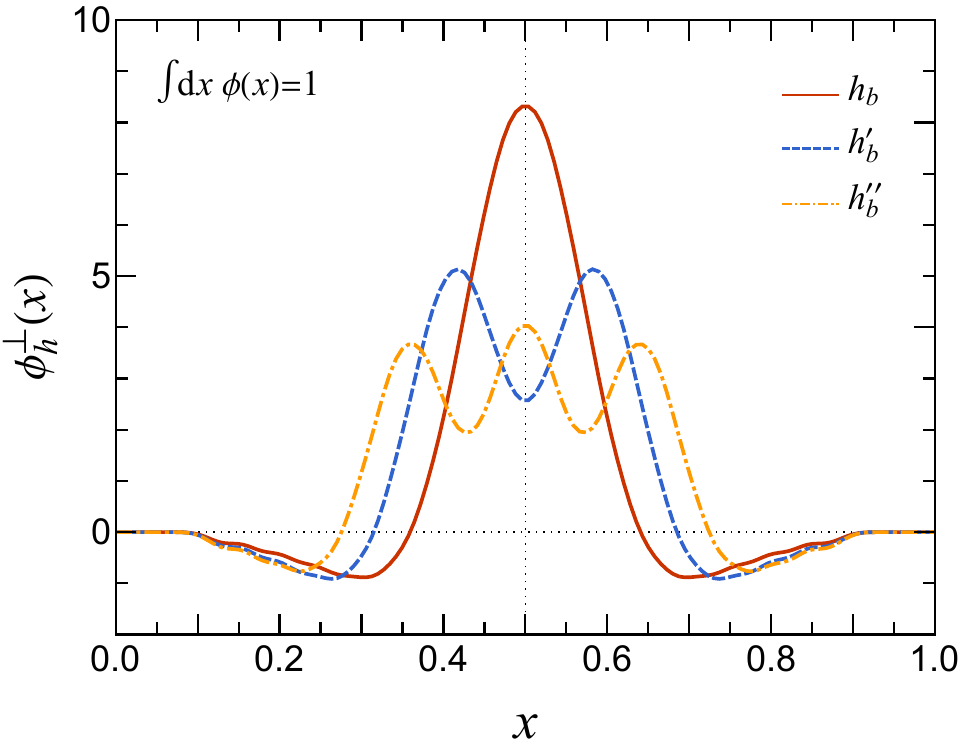}    
    \caption{Leading-twist distribution amplitudes of the $P$-wave axial vector quarkonia ($1^{+-}$) $h_c, h_b$ and their radial excitations.}
    \label{fig:h_DAs}
\end{figure}

The LCDAs for the $P$-wave axial vector quarkonia $\chi_{c1}$ and $\chi_{b1}$ (including both ground states and radial excitations) are presented in Fig.~\ref{fig:A_DAs}. A striking feature of these distributions is the emergence of a novel ``W"-shaped structure in the ground states. This profile originates directly from the relativistically induced $S/D$ partial-wave components ($\psi_{\uparrow\downarrow-\downarrow\uparrow}^{(m_j=0)}$) within their LFWFs, a feature fundamentally absent in non-relativistic models. Structurally, the invariant functions $\phi_{2,3,6}$ exhibit $S$-wave characteristics, whereas $\phi_{1,4}$ drive $D$-wave behaviors. The resulting ``W"-shaped profile is a direct manifestation of the interference between these $S$- and $D$-wave components. Such non-trivial distributions provide clear quantitative signatures of relativistic dynamics that can, in principle, be probed in hard exclusive processes. 
Furthermore, as illustrated in Fig.~\ref{fig:h_DAs}, this dynamical partial-wave mixing is not isolated to the $\textsf{C}=+1$ sector; the transverse LCDAs of the $\textsf{C}=-1$ axial vectors $h_c$ and $h_b$ exhibit precisely analogous ``W"-shaped interference patterns, underscoring the universality and robustness of these relativistic structural effects across the axial vector family.

\begin{figure}
    \centering
    \subfigure[\ charmonia $\phi_\text{even}(x)$]{\includegraphics[width=0.48\textwidth]{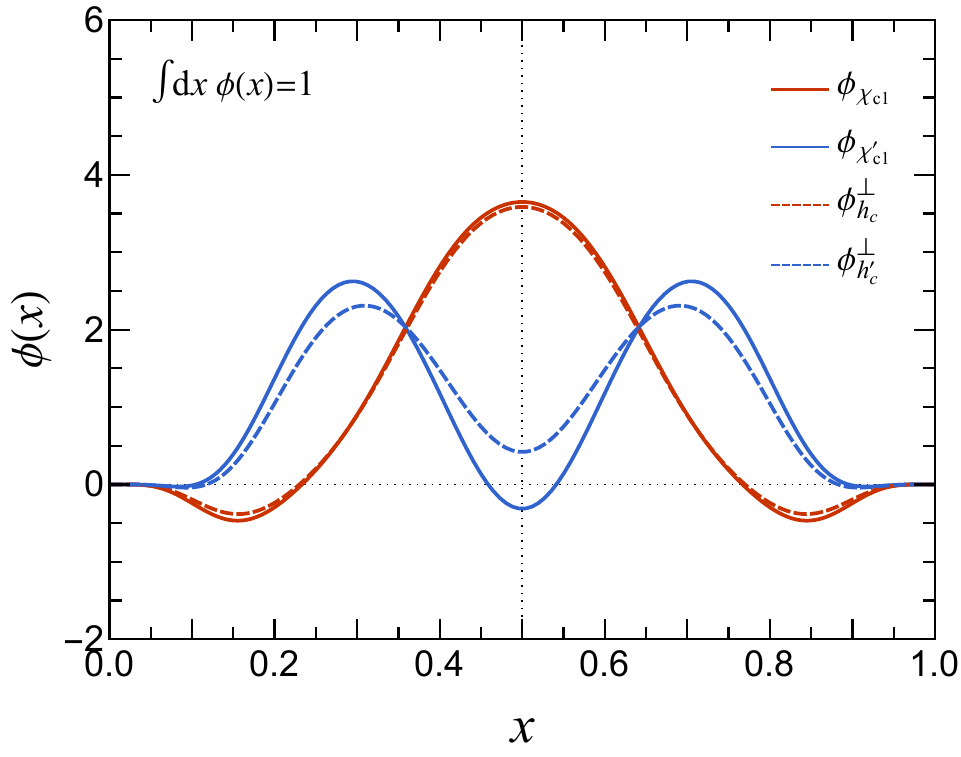}}\quad
    \subfigure[\ charmonia $\phi_\text{odd}(x)$]{\includegraphics[width=0.48\textwidth]{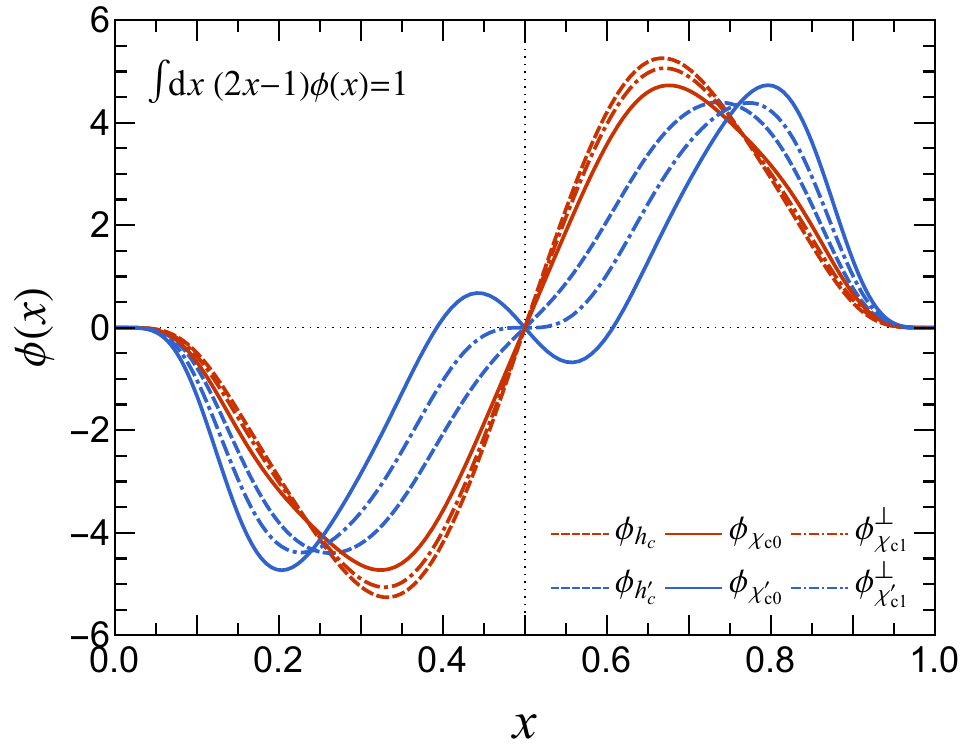}}
    \subfigure[\ bottomonia $\phi_\text{even}(x)$]{\includegraphics[width=0.48\textwidth]{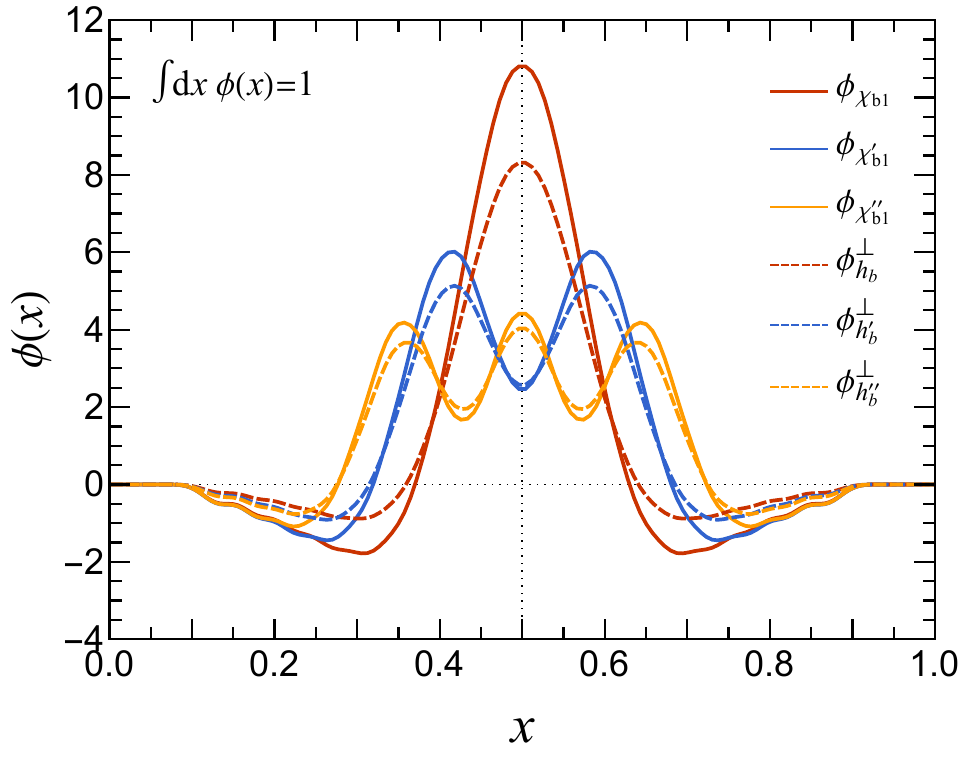}}\quad
    \subfigure[\ bottomonia $\phi_\text{odd}(x)$]{\includegraphics[width=0.48\textwidth]{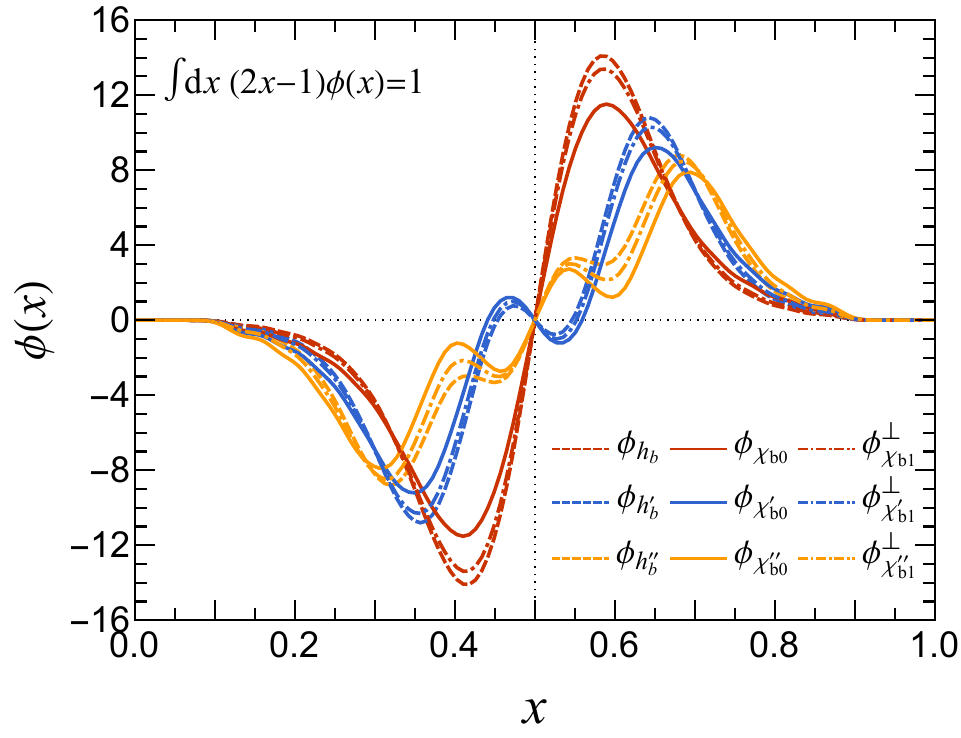}}
    \caption{Comparison of $P$-wave quarkonia distribution amplitudes with even and odd parities in $x$.}
    \label{fig:DAs_even_odd}
\end{figure}

In the non-relativistic limit, the various $P$-wave LCDAs simplify into two distinct functions defined by their longitudinal parity behavior \cite{Braguta:2008qe}: an antisymmetric (odd) set, $\phi_\text{odd} = \phi_{S} = \phi_{A}^\perp = \phi_{h}$, and a symmetric (even) set, $\phi_\text{even} = \phi_A = \phi_h^\perp$. The BLFQ results evaluated across this regime are displayed in Fig.~\ref{fig:DAs_even_odd}. As the constituent quark mass increases from charmonium to bottomonium, the computed LCDAs systematically converge toward these non-relativistic predictions.

\begin{figure}
    \centering
    \includegraphics[width=0.45\textwidth]{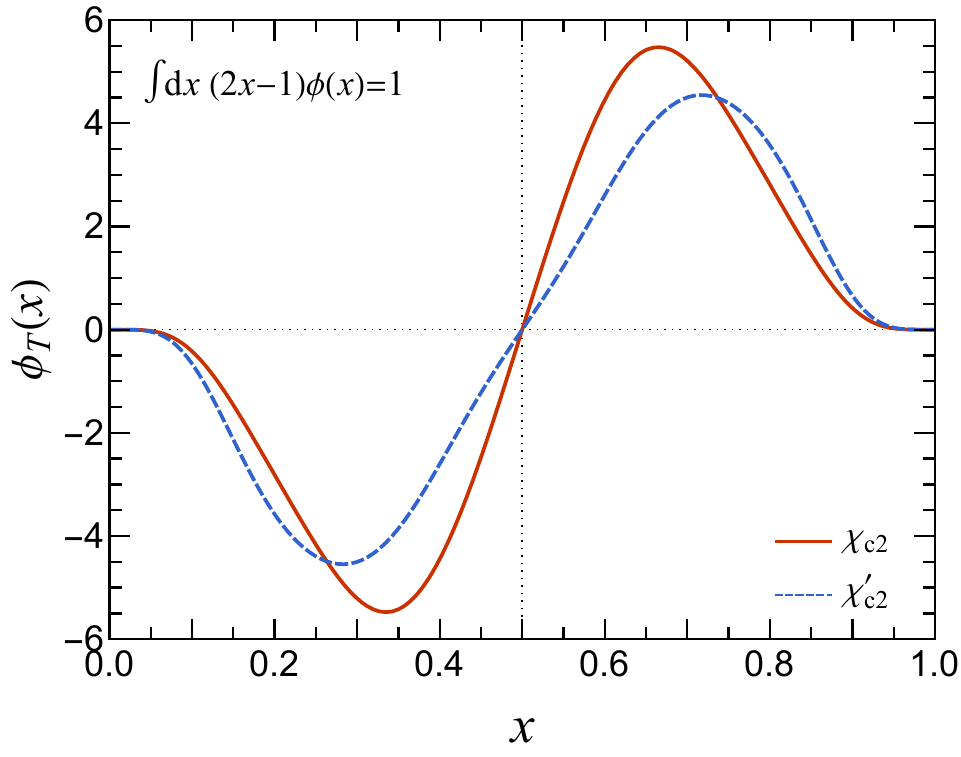}
    \includegraphics[width=0.46\textwidth]{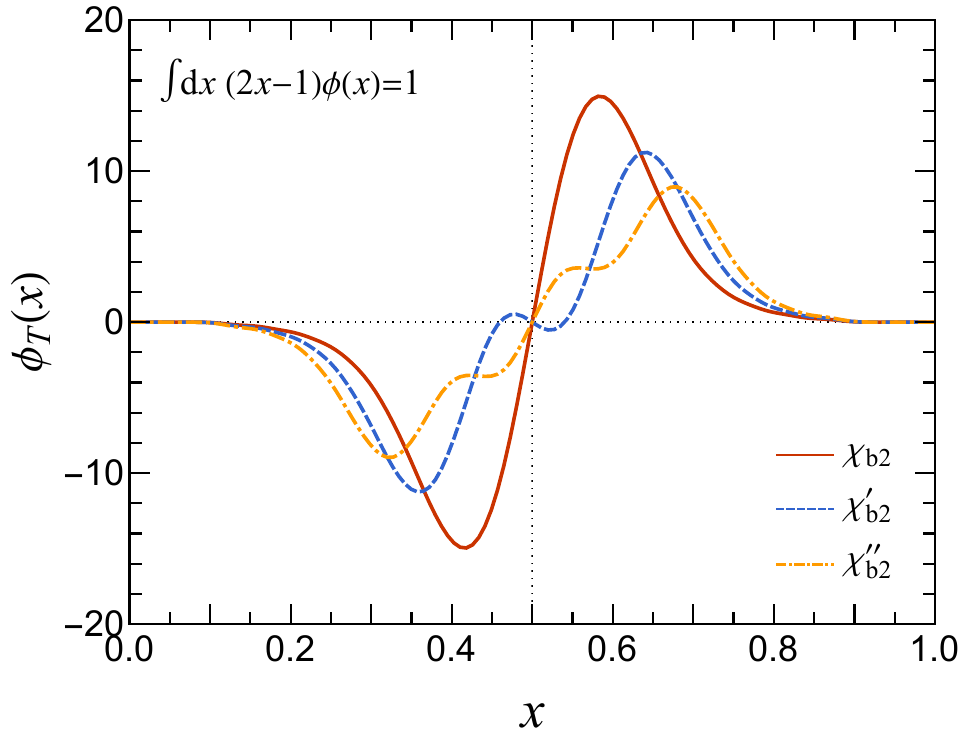}
    \includegraphics[width=0.45\textwidth]{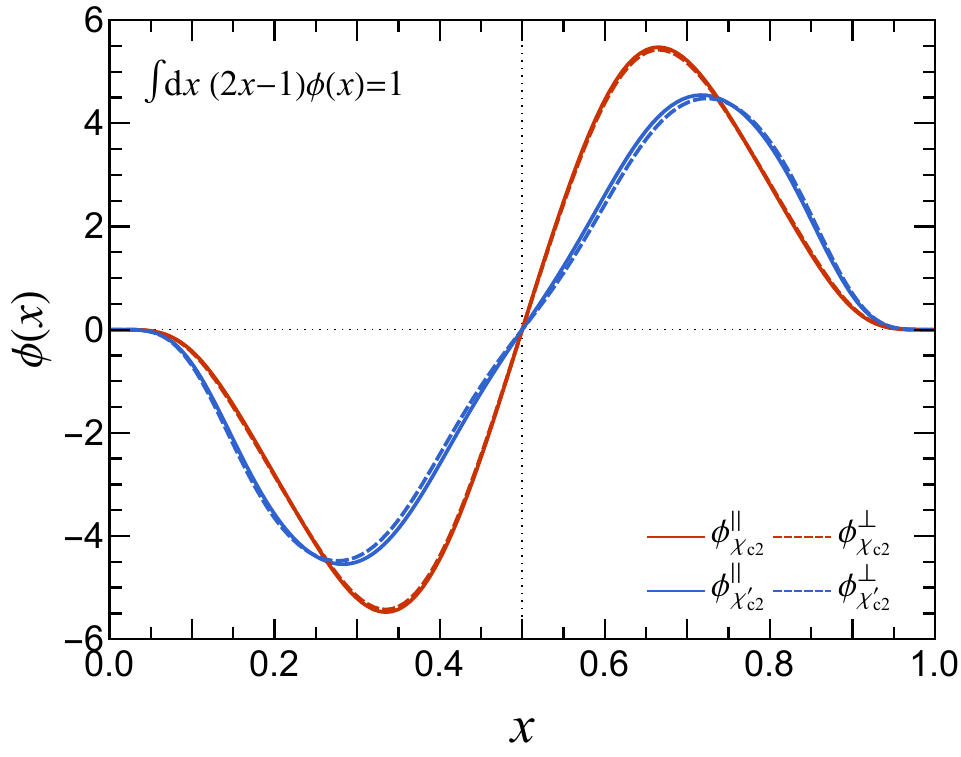}
    \includegraphics[width=0.46\textwidth]{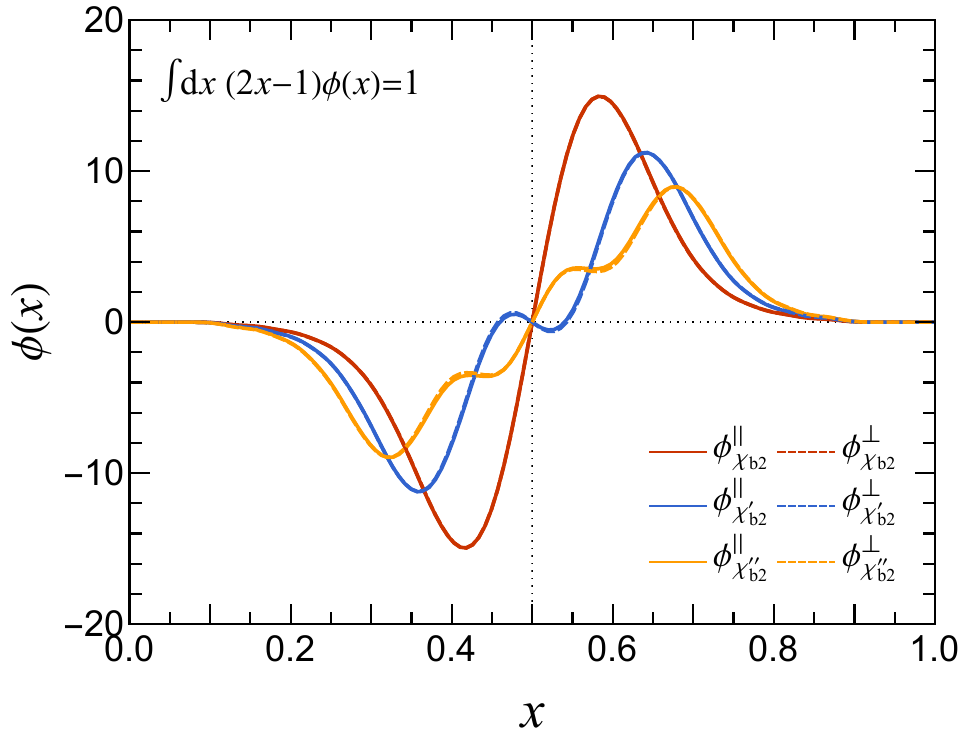}
    \caption{Distribution amplitudes of the $P$-wave tensor quarkonia $\chi_{c2}, \chi_{b2}$ and their radial excitations.}
    \label{fig:T_DAs}
\end{figure}

Figure~\ref{fig:T_DAs} presents the longitudinal and transverse LCDAs for the $2^{++}$ tensor quarkonia. Both amplitudes are strictly antisymmetric under longitudinal momentum exchange ($x \to 1-x$). In the heavy quark limit, these distinct longitudinal and transverse distributions converge to the same universal function, perfectly aligning with the expectations of the non-relativistic quark model.

\section{Summary}\label{sec:summary}

In this work, we investigated the leading-twist light-cone distribution amplitudes (LCDAs) of $P$-wave quarkonia, specifically the scalar ($\chi_0$), axial vector ($\chi_1$ and $h$), and tensor ($\chi_2$) states, within the basis light-front quantization (BLFQ) framework. The underlying effective interaction incorporates both light-front holographic confinement and short-distance one-gluon exchange dynamics.
Within the light-front wave function (LFWF) representation, we demonstrated that the leading-twist LCDAs of these different mesons admit a unified formalism: they are all directly related to their respective wave function components with zero orbital angular momentum projection ($L_z = 0$).

The overall shapes of the LFWFs are fundamentally constrained by discrete and continuous symmetries, including Lorentz boosts, rotational invariance along $z$-axis, mirror parity, and charge conjugation. Incorporating all these symmetries, we gave the most general structures of the quarkonia LFWFs in the valence sector.  
For heavy systems such as charmonia and bottomonia, we showed that the LFWFs properly approach their expected non-relativistic reductions as the quark mass increases. Nevertheless, our results emphasize that relativistic effects still play a crucial role in shaping the internal structure of these hadrons. Relativistic dynamics naturally generate mixed partial waves beyond the configurations predicted by the traditional non-relativistic constituent quark model. For instance, we observe robust $S/D$-wave components in the axial vector mesons. These exotic, relativistically induced partial waves directly manifest themselves in the behavior of the LCDAs. A notable example is the axial vector meson $\chi_1$, where the $S/D$-wave spin-singlet LFWF $\psi_{\uparrow\downarrow-\downarrow\uparrow/A}^{(m_j=0)}(x, \vec k_\perp)$ gives rise to a distinct ``W"-shaped distribution amplitude.

Looking forward, the LCDAs obtained in this study provide essential non-perturbative inputs for phenomenological studies of modern high-energy scattering experiments. In particular, they will be instrumental for theoretical descriptions of hard exclusive processes, such as the production of $P$-wave quarkonia at current and future high-luminosity colliders. A natural extension of this work will involve applying these wave functions to calculate transition form factors and higher-twist distributions, further bridging the gap between fundamental dynamics and experimental observables.

\acknowledgements

We wish to thank W. Qian for fruitful discussions. 
This work is supported in part by the Chinese Academy of Sciences under Grant No.~YSBR-101.
M.L. is supported by new faculty startup funding by Huazhong University of Science and Technology.

\end{document}